\documentstyle[psfig,eqsecnum,prd,aps,epsf]{revtex}
\begin{document}

\input epsf
\draft

\title{
Boson stars driven to the brink of black hole formation
}

\author{Scott H. Hawley${}^{1}$  
  \thanks{ E-mail address: shawley\,@\,physics.utexas.edu} 
and
Matthew W. Choptuik${}^{1,2}$    
  \thanks{ E-mail address: choptuik\,@\,physics.ubc.ca}
}
\address{${}^1$Center for Relativity, Department of Physics, 
               University of Texas at Austin, Austin, TX 78712-1081 USA}
\address{${}^2$ 
	 CIAR Gravity and Cosmology Program, 
         Department of Physics and Astronomy, 
         University of British Columbia, Vancouver, British Columbia, 
         Canada V6T 1Z1
}

\maketitle

\begin{abstract}
We present a study of black hole threshold phenomena for a
self-gravitating, massive complex scalar field in spherical symmetry.
We construct Type I critical solutions dynamically by tuning a
one-parameter family of initial data consisting of a boson star and a
massless real scalar field.  The massless field is used to perturb the
boson star via a  purely gravitational interaction which results in a
{\em significant} transfer of energy from the massless field to the
massive field.  The resulting (unstable) critical solutions, which
display great similarity with unstable boson stars, persist for a
finite time before either dispersing most of the mass to infinity
(leaving a diffuse remnant) or forming a black hole.  To further the
comparison between our critical solutions and boson stars, we verify
and extend the linear stability analysis of Gleiser and Watkins [M.
Gleiser and R.  Watkins, Nucl. Phys. B{\bf 319} 733 (1989)] by
providing a method for calculating the radial dependence of boson star
quasinormal modes of nonzero frequency.  The frequencies observed in our
critical solutions coincide with the mode frequencies obtained from
perturbation theory, as do the radial profiles of many of the
modes.  For critical solutions less than $90\%$ of the maximum boson star 
mass $ M_{\rm max} \simeq 0.633 M_{Pl}^2/m$, the existence of a small halo of 
matter in the tail of the solution distorts the profiles which
otherwise agree very well with unstable boson stars.   These halos
appear to be artifacts of the collision between the original boson star
and the massless field, and do not appear to belong to the true
critical solutions, which are interior to the halos and which do in
fact correspond to unstable boson stars.  It appears that unstable
boson stars are unstable to dispersal (``explosion'') in addition to
black hole formation, and given the similarities in macroscopic
stability between boson stars and neutron stars, we suggest that those
neutron star configurations at or beyond the point of instability may
likewise be unstable to explosion.  
\end{abstract}


\section{Introduction}
Over the past decade, detailed studies of models of gravitational
collapse have revealed that the threshold of black hole formation is
generically characterized by special, ``critical'' solutions.  The
features of these solutions are known as ``critical phenomena,'' and
arise in even the simplest collapse models, such as a model consisting
of a single, real, massless scalar field, minimally coupled to the
general relativistic field in spherical symmetry~\cite{Mattcrit}.
Although we present a brief overview of black hole critical
phenomena here, we suggest that interested readers consult the excellent
reviews by Gundlach \cite{CarstenRev,CarstenLivRev} for many additional
details.

Black hole critical solutions can be constructed dynamically via
simulation, {\it i.e.} via solution of the full time-dependent PDEs
describing the particular model, by considering one-parameter families
of initial data which are required to have the following
``interpolating'' property: for sufficiently large values of the family
parameter, $p$, the evolved data describes a spacetime containing a
black hole, whereas for sufficiently small values of $p$, the
matter-energy in the spacetime disperses to large radii at late times,
and {\em no} black hole forms.  For any such family, there will exist a
critical parameter value, $p=p^\star$, which demarks the onset, or threshold, 
of black hole formation.  To date at least, it has invariably turned out that 
the solutions which appear in the strongly-coupled regime of the
calculations (i.e. the critical solution), are almost totally 
{\em independent} of the
specifics of the particular family used as a generator.   In fact, the only 
initial-data dependence which has been observed so far in critical collapse 
occurs in models for which there is more than one distinct black-hole-threshold
solution.  In this sense
then, black hole critical solutions are akin to, for example, the
Schwarzschild solution, which can be formed through the collapse of
virtually any type and/or shape of spherically distributed matter.  In
particular, like the Schwarzschild solution, black hole critical
solutions possess additional symmetry (beyond spherical symmetry)
which, to date, has either been a time-translation symmetry, in which the
critical solution is static or periodic, or a scale-translation
symmetry (hometheticity), in which the critical solution is either
continuously or discretely self-similar (CSS or DSS).

However, in clear contrast to the Schwarzschild solution, black hole
threshold solutions are, by construction, {\em unstable}.  Indeed, after
seminal work by Evans and Coleman \cite{EC} and by Koike et al
\cite{Koike}, we have come to understand that critical solutions are in
some sense {\em minimally} unstable, in that they tend to have
precisely one unstable mode in linear perturbation theory.  Thus
letting $p\rightarrow p^{\star}$ amounts to minimizing or ``tuning away''
the initial amplitude of the unstable mode present in the system.

As already suggested, two principal
types of critical behavior have been seen in black hole threshold studies;
which type is observed depends, in general, upon the type of matter model {\em
 and}
the initial data used---as mentioned, some models exhibit both 
types of behavior.   Historically, one of us termed these Type I and Type II 
solutions,
in a loose analogy to phase transitions in statistical mechanics, but 
at least at this juncture, we could equally well label the critical
solutions by their symmetries (i.e. static/periodic or CSS/DSS) .  For
Type I solutions, there is a finite minimum black hole mass which can
be formed, and, in accord with their static/periodic nature, there is a scaling
 
law,  $\tau \sim -\gamma \ln|p-p^{\star}|$, relating the lifetime, $\tau$, 
of a near-critical solution to the proximity of the solution to the 
critical point.  Here $\gamma$ is a model-specific exponent which is 
the reciprocal of the real part of the eigenvalue associated with the 
unstable mode.
On the other hand, Type II critical behavior---less relevant to the 
current study---is characterized by 
arbitrarily small black hole mass at threshold, and 
critical solutions which are generically {\em self-similar}. 

The direct construction, or simulation, of critical solutions, has thus 
far been performed almost exclusively within the {\em ansatz} of
spherical symmetry.  In the spherical case one {\em must} couple to at
least one matter field for non-trivial dynamics, and spherically
symmetric critical solutions for a considerable variety of models have
now been constructed and analyzed.  In addition to the massless scalar
case mentioned above, these include solutions containing a perfect
fluid \cite{EC,Dave}, a scalar non-Abelian gauge field\cite{MattYM},
and particularly germane to the current work, a {\em massive} real
scalar field \cite{Brady}.  The work of Abrahams and Evans\cite{AE},
which considered {\em axisymmetric} critical collapse of gravitational
waves, remains notable for being the only instance 
of a reasonably well-resolved non-spherical critical solution
\cite{3DCritNote}.

Our current interest is a critical-phenomena-inspired study of the
dynamics associated with ``boson stars'' \cite{Kaup,RB,Colpi}, a class
of equilibrium solutions to the Einstein-Klein Gordon system for
massive complex fields, which are supported against gravitational
collapse by the effective pressure due to the dispersive nature of a
massive Klein-Gordon field.  (For extensive reviews on the subject of
boson stars, see Jetzer \cite{Jetzer} or Mielke and Schunck
\cite{Mielke}.) We know from the studies by Gleiser and Watkins
\cite{GW} and by Lee and Pang \cite{Lee}, that there exists a critical
value of the central density which marks the transition between boson
stars which are stable with respect to infinitesimal radial
perturbations, and those which are unstable.  The dynamical simulations
of Seidel and Suen \cite{SandS} revealed scenarios in which a boson
star on the unstable branch would either form a black hole or radiate
scalar material and form a boson star on the stable branch.  Their
study is extended in this paper, in which we consider dynamical changes
to the geometry of a boson star which are large enough to bring it to
the threshold of black hole formation.

As already mentioned, another paper closely related to this work is that of 
Brady {\it et al.} \cite{Brady}, which described a dynamical study of critical
solutions of a massive real scalar field.  Those authors demonstrated 
scenarios in which black holes could be formed with arbitrarily small 
mass (Type II
transitions), {\em and} those in which the black holes formed had a finite
minimum mass (Type I transitions).  The boundary between these regimes
seemed to be the relative length scale of the pulse of initial data
compared to the Compton wavelength associated with the boson mass.
Initial data which was ``kinetic energy dominated'' evolved in a manner
essentially similar to the evolution of a massless scalar field.
Initial data pulses having widths larger than the length scale set by
the boson mass were ``potential dominated,'' providing a characteristic
scale for the formation of the critical solutions.  Brady {\it et al.}
found that the resulting Type I critical solutions corresponded to a
class of equilibrium solutions discovered by Seidel and Suen
\cite{OSS}, which are called ``oscillating soliton stars.'' These
soliton stars share many characteristic with the complex-valued boson
stars, such as the relationship between the radius and mass of the
star.  Both types of ``stars'' have a maximum mass, and show the same
overall behavior as ``real'' (fermion) stars in terms of the turn-over
in their respective stability curves.  Interestingly, although the soliton
 stars are not
static---they are periodic (or quasi-periodic)---many of the same macroscopic
 properties 
seen in fluid stars are still observed.

In this paper, we construct critical solutions of the Einstein
equations coupled to a massive, {\em complex} scalar field
dynamically, by
simulating the implosion of a spherical shell of {\em massless}
real scalar field around an ``enclosed'' boson star.  The massless
field implodes toward the boson star and the two fields undergo a
(purely gravitational) ``collision.'' The massless pulse then passes
through the origin, explodes and continues to $r\rightarrow \infty$,
while the massive complex (boson star) field is compressed into a state
which ultimately either forms a black hole or disperses.  We can thus
play the ``interpolation game'' using initial data which result in
black hole formation, and initial data which give rise to dispersal:
specifically, we vary the initial amplitude of the {\em massless} pulse
to tune to a critical solution.  We analyze the black hole threshold
solutions obtained in this manner, and discuss the similarities between
our critical solutions for the self-gravitating complex massive scalar 
field and boson stars
on the unstable branch.  To further this discussion, we extend the work
of Gleiser and Watkins \cite{GW} and compare the results of the
simulations with those of linear perturbation theory.

The layout of the remainder this paper is as follows:  In Section II,
we describe the mathematical basis for our numerical simulations.  In
Section III, we present results from our simulations, in which the Type
I character of the critical solutions is demonstrated, along with the
close similarities one finds between the features of the critical
solutions and those of boson stars.  In most of the critical solutions
we find a halo of mass near the outer edge of
the solution which is not a feature of boson star equilibrium data.
Inside this halo, however, the critical solutions match the
boson star profiles very well.  In Section IV, we give a synopsis of our linear
stability analysis of boson star quasinormal modes, from which we
obtain the boson star mode frequencies as functions of the central
value of the modulus of the complex field.  Section V concerns the
radial profiles of the perturbative modes {\em per se}, and includes a
comparison of the mode shapes and frequencies obtained from
perturbation theory with our simulation data.  The modes obtained by
these two different methods agree well with each other, although the
additional oscillatory mode in our simulation data is only shown to
agree with the corresponding boson star mode in terms of the
oscillations in the metric and not in the field (possibly as an
artifact of our simplistic approach to extracting this mode from the
simulation).
In Section V we provide further discussion regarding the properties of
the halos surrounding the critical solutions.

Conclusions in Section VI are followed by appendices giving tables of
mode frequencies versus the central field value of the boson star,
details about our finite difference code, and details of our linear
stability analysis, which includes a description of our algorithm for
finding the frequencies of boson star modes.

\section{Scalar Field Model}
A boson star is described by a complex massive scalar field
$\phi$, minimally coupled to gravity as given by general
relativity.  We work solely within the context of classical field
theory, and choose units in which $G$ and $c$ are unity.
Furthermore, since all lengths in the problem can be scaled by 
the boson mass $m$ \cite{Colpi}, we choose $m=1$.
To the usual boson star model, we add an additional,
massless real scalar field, $\phi_3$, which is also 
minimally coupled to gravity.
This additional scalar field will be used to dynamically ``perturb'' the boson
 star.

The equations of motion for the system are then the Einstein equation and
Klein-Gordon equations:

\begin{equation} G_{ab} = R_{ab}- {1\over 2}g_{ab}R = 8\pi 
    \left( T^C_{ab}(\phi) + T^R_{ab}(\phi_3) \right) \end{equation} 
\begin{equation} \Box\phi - m^2\phi = 0
\label{firstkg}\end{equation}
\begin{equation} \Box\phi_3  = 0\end{equation}
where
\begin{equation}
8\pi T^C_{ab}(\phi)  = \partial_a\phi\partial_b\phi^{*} + \partial_a\phi^{
*}\partial_b\phi
		  - g_{ab}\left(\partial_c\phi\partial^c\phi^{*} + m^2|\phi|^2 \right)
, \end{equation}
\begin{equation}
8\pi T^R_{ab}(\phi_3) = 2 \partial_a \phi_3 \partial_b \phi_3 - 
   \, g_{ab} \partial^c \phi_3 \partial_c \phi_3
, \end{equation}
and $\Box$ is the D'Alembertian operator.  While more general potentials in
 (\ref{firstkg}) 
have been employed recently \cite{BSS,SchunckTorres}, we will restrict our
 discussion
to the simplest case, {\it i.e.}  merely the $m^2\phi^2$ potential.  We 
also stress that the complex scalar field, $\phi$, and the massless,
real scalar field, $\phi_3$ are coupled {\em only} through gravity---in 
particular we do not include any interaction potential $V_I(\phi,\phi_3)$.

Restricting our attention to spherical symmetry, we write the most general
spherically-symmetric metric using Schwarzschild-like ``polar-areal"
 coordinates
\begin{equation} ds^2 = -\alpha^2(t,r) dt^2 + a^2(t,r) dr^2 + r^2 d\Omega^2 \,
 ,
\label{metric}
\end{equation}
and generally make use of the ``3+1'' formalism of Arnowitt, Deser and Misner 
\cite{ADM} which regards spacetime as a foliation of spacelike hypersurfaces 
parameterized by $t$.

We write the (spherically-symmetric) complex field, $\phi(t,r)$, in terms of
 its components
\begin{equation}\phi(t,r) = \phi_1(t,r) + i \phi_2(t,r)
\label{phidecomp} \end{equation} 
where $\phi_1(t,r)$ and $\phi_2(t,r)$ are each real.
Again, since our model includes no self-interaction (anharmonic) potential
for the complex field, $\phi_1$ and $\phi_2$ are only coupled 
through the gravitational field.

We then define 
\begin{equation}\Phi_1(t,r) \equiv {\phi_1'} \ \ \ \ \ \ \ \ \Phi_2(t,r) \equiv
 {\phi_2'} 
\label{defPhi}\end{equation} 
\begin{equation}\Pi_1(t,r) \equiv {a\over\alpha}\dot{\phi_1} \ \ \ \ \ \ \ \ 
		\Pi_2(t,r) \equiv {a\over\alpha}\dot{\phi_2}, \label{defPi}\end{equation}
\begin{equation} \Phi_3(t,r) = {\phi_3'} \ \ \ \ \ \ \ \Pi_3(t,r) =
 {a\over\alpha}\dot{\phi_3}. \end{equation}
where $'\equiv \partial/\partial r$ and $\dot{}\equiv \partial / \partial t.$

With these definitions, the equations we solve are
the Hamiltonian constraint,
\begin{equation}
 {a'\over a}  =  {1-a^2\over 2r} + {r\over 2} \left[ 
			\Pi_1{}^2 + \Pi_2{}^2 + \Pi_3{}^2 
		  + \Phi_1{}^2 + \Phi_2{}^2 + \Phi_3{}^2 
	 +  a\left(\phi_1{}^2 + \phi_2{}^2\right) \right],
\label{Ham}
\end{equation}

(where $\Pi_1{}^2$ should be read as $(\Pi_1)^2$),
the slicing condition,
\begin{equation} {\alpha'\over\alpha} = {a^2 - 1 \over r} + {a' \over a} 
			 - a^2 r  (\phi_1{}^2 + \phi_2{}^2),
\label{Slice}\end{equation}

and the Klein-Gordon equations,
\begin{equation}
\dot{\Pi}_k = 3 {\partial \over \partial r^3} \left( {r^2 \alpha \over a}
\Phi_k \right)' -  \alpha a \phi_k( 1 - \delta_{3k}) ,
\label{KG} 
\end{equation}
where $k=1,2,3$ and $\delta_{3k}$ is a Kronecker delta used to denote
the fact that $\phi_3$ is a massless field.

We also have equations which are used to update the spatial gradients of the
scalar fields, as well as the scalar fields themselves.  These follow directly
from the definitions
(\ref{defPhi}) and (\ref{defPi}):
\begin{equation}
 \dot{\Phi}_k  =  \left( {\alpha\over a}\Pi_k \right)'
  \label{PhiPirel}
\end{equation}
\begin{equation}
\phi_k = \int_0^r \Phi_k d\tilde{r}
\label{phi_int}
\end{equation}
Equations (\ref{Ham})--(\ref{phi_int}) are solved numerically using
the second order finite difference method described in Appendix B.

Initial conditions for our simulations are set up as follows.
First, initial data for the massive field are constructed from the boson star
{\em ansatz}
\begin{equation}
\phi(t,r) = \phi_0(r)e^{-i\omega t},
\label{BSansatz}
\end{equation}
 where we let $\phi_0(r)$ be real.  
Substitution of this {\em ansatz} into the full set of equations
(\ref{Ham})-(\ref{phi_int}), 
yields a system of ordinary
differential equations (ODEs), whose solution, for a given value of the central
 field
modulus,  is found by ``shooting,'' as described in \cite{RB}.  Once the boson 
star data is in hand, we add the perturbing massless field by freely specifying
$\Phi_3$ and $\Pi_3$.  At this point, all matter quantities have been specified
;
the initial geometry, $a(0,r)$ and $\alpha(0,r)$ is then fixed by the
 constraint and slicing 
equations~(\ref{Ham}) and (\ref{Slice}).

In relating the simulation results which follow, it is useful to consider the
individual contributions of  the complex and real fields to the
total mass distribution of the space-time, in order that we can meaningfully 
and unambiguously discuss, for example, the exchange of mass-energy from the 
real, massless field to the massive, complex field.
By Birchoff's theorem, in any vacuum region, the mass enclosed by a sphere of
 radius $r$ at a 
given time $t$ is given by the Schwarzschild-like mass aspect function 
$M(t,r) = r(1-1/a^2)/2$.  However, at locations occupied by matter, $M(t,r)$
cannot necessarily be usefully interpreted as a ``physical'' mass.
In polar-areal coordinates, the mass aspect function is related to the local
 energy 
density $\rho(t,r)$ by
\begin{equation} {\partial M(t,r)\over \partial r} = r^2 \rho(t,r), 
\label{massaspect}
\end{equation} 
with $\rho(t,r)$ given in our case by
\begin{equation}
 \rho(t,r) = {1\over 2 a^2}\left[
	  \Pi_1{}^2+\Pi_2{}^2 + \Phi_1{}^2+\Phi_2{}^2
		 +  {a^2}\left( \phi_1{}^2 + \phi_2{}^2 \right) \right]
		 + {1\over 2 a^2}\left[ \Pi_3{}^2 + +\Phi_3{}^2 \right] \, .
\label{rhoeq} 
\end{equation}
Here, we have explicitly separated the contributions from the complex and
real fields.
Since $\partial M/\partial r$ is given by a linear combination
of the contributions from each field, we can decompose
$\partial M/\partial r$ so that,
in instances where there is no overlap in the supports of the 
distinct fields, we can unambiguously 
refer to the mass due to one field or the other.  That is,
we can refer to the individual contributions of each field to
the total mass as being physically meaningful masses in their
own rights.
Then, by integrating the contribution of {\em each} field to $\partial M 
/\partial r$ 
over some range of radius $(r_{\rm min} \cdots r_{\rm max})$, 
(where there is {\em some} region of vacuum 
starting at $r_{\rm min}$ and extending inward, and {\em some} region of vacuum
starting at $r >= r_{\rm max}$ and extending outward), and demanding that 
none of the other type of field is present
in the domain of integration, we can obtain
a measure of the mass due to each field.

Motivated by such considerations, we define an energy density for the complex
 field, $\rho_C$, as
\begin{equation} 
 \rho_C(t,r) = {1\over 2 a^2}\left[
			 \Pi_1{}^2+\Pi_2{}^2 + \Phi_1{}^2+\Phi_2{}^2 
		 +  {a^2}\left( \phi_1{}^2 + \phi_2{}^2 \right) \right]
,
\end{equation}
with a corresponding mass aspect function, $M_C(t,r)$, given by
 \begin{equation} M_C(t,r) = \int_0^r \tilde{r}^2 \,\rho_C\, d\tilde{r}\, . 
\label{M_C}
\end{equation}
Similarly, the energy density due to the real field is defined as
\begin{equation}
 \rho_R(t,r) \equiv {1\over 2 a^2}\left[ \Pi_3{}^2 + +\Phi_3{}^2 \right],
\end{equation}
with a corresponding mass aspect function, $M_R(t,r)$ given by
$$
  M_R(t,r) = \int_0^r \tilde{r}^2\, \rho_R\, d\tilde{r}. 
$$

We again emphasize that
in regions where the supports of the different fields overlap
(and in non-vacuum regions in general)
it may not be possible to ascribe physical 
meaning to the individual mass aspect functions defined above.
(However, even in such instances,  these functions are still useful 
diagnostics.)
Most importantly, where the supports of the fields {\em do} overlap,
and only in these regions, it is possible for mass-energy to be exchanged from
 one scalar
field to the other---{\em through the gravitational field}---while the sum $M_C
 + M_R  = M$ (measured in an exterior
vacuum region) is conserved.
The quantities given above allow us to measure this
exchange of mass by looking at the profiles $M_C(t,r)$ and $M_R(t,r)$ before
 and after a time 
when the fields are interacting.  This is shown in the next section.

As a further consideration, we point out that
the $U(1)$ symmetry of the complex field implies that there
is a conserved Noether current, $J^\mu$, given by
\begin{equation}
  J^\mu = {i\over 8\pi}g^{\mu\nu}(\phi\partial_\nu\phi^* - \phi^
*\partial_\nu\phi).
\end{equation}
The corresponding conserved charge or ``particle number'' $N$ is 
$$ N = \int_0^\infty r^2 \sqrt{-g} J^t.$$
We may also wish to regard $N$ as a function of $t$ and $r$ by 
integrating the above function from zero to some finite radius, in which case 
\begin{equation}
    {\partial N(t,r)\over \partial r} = r^2 \left(\Pi_1\phi_2 - \Pi_2
 \phi_1\right).
\end{equation}

Some authors have considered the difference $M_C - mN$ to be a sort of 
``binding energy'' of the complex field \cite{Jetzer}, however 
this quantity does not correspond to any transition in the stability of boson 
stars, and we have not found it to be very useful in understanding the dynamics
 
of our simulations.

Finally, following Seidel and Suen \cite{SandS}, we define a radius 
$R_{95}(t,r)$ for
the boson star implicitly by $M_C|_{R_{95}} = 0.95\, M_C|_{r\rightarrow\infty}$
.
Alternatively, we will also consider a radius $R_{63}(t,r)$ which encloses
$(1-e^{-1})\sim 63\%$ of $M_C|_{r\rightarrow\infty}$, and which will
include the ``bulk'' of a boson star but will neglect the ``tail''.

\section{Simulation Results}
We choose the initial data for the complex field to be a boson star at the 
origin, with a given central density $\phi_0(0)$. 
For the massless field $\phi_3(0,r)$, we choose one of the families in 
Table \ref{tab:families}.
We generate critical solutions by tuning the amplitude $A$ of $\phi_3(0,r)$
(holding the position $r_0$ and width $\Delta$ constant) 
using a bisection
search, until the resulting solution is arbitrarily close (i.e. within
some specified precision) to the point of unstable equilibrium between
dispersal and black hole formation.

\begin{figure}
\epsfxsize = 12cm
\epsfysize = 12cm                                            
\centerline{\epsffile{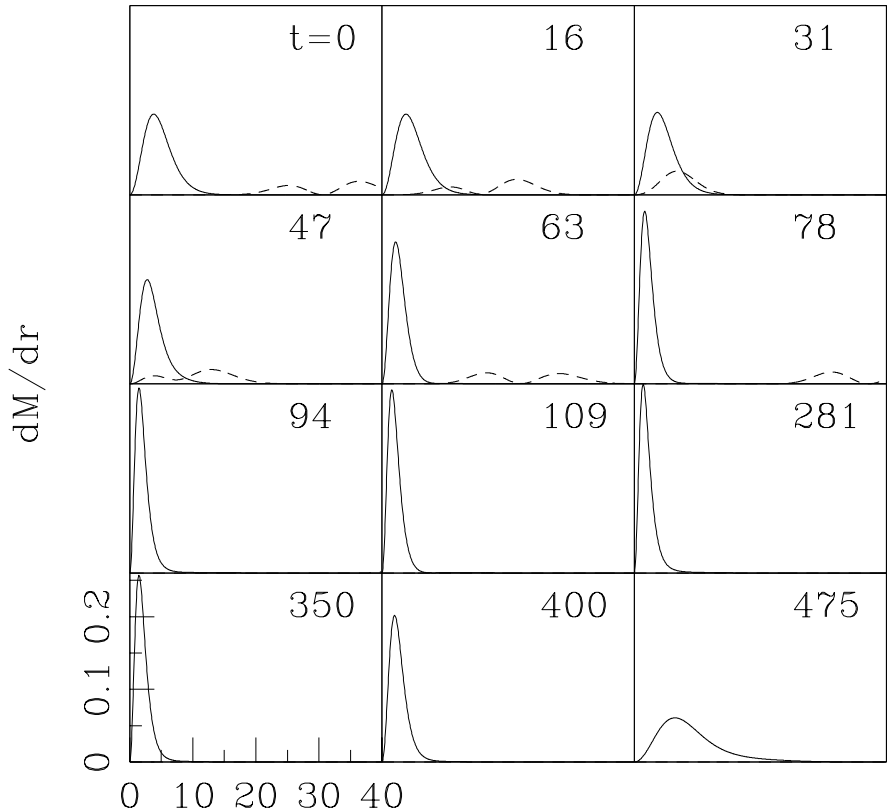}}
\caption{
Evolution of a perturbed boson star with
$\phi_0(0)=0.04\times\sqrt{4\pi}$ and mass $M_C=0.59 M_{Pl}^2/m$.
This shows contributions to $\partial M/\partial r$ due to the massive field
 (solid
line) and massless field (dashed line).  We start with a stable boson
star centered at the origin, and a pulse of massless field given by 
Family $\rm I$ with $r_0=30$ and $\Delta=8$.  (We see two peaks in
$dM/dr$ of massless field because it is only the gradients of $\phi_3$, 
not $\phi_3$ itself, which contribute to $M_R(r,t)$ for a massless field.)
In the evolution shown above, the
pulse of massless field enters the region containing the bulk of the boson star
($t\simeq 15$), implodes through the origin ($t \simeq 30$) and leaves
the region of the boson star ($t \simeq 50$).  Shortly after the
massless pulse passes through the origin, the boson star collapses into
a more compact configuration, about which it oscillates for a long time
before either forming a black hole or dispersing.  (The case of
dispersal is shown here.) 
 Note that the perturbing field $\phi_3$
passes through the boson star and exits the region containing most of
the star, even before the massive field reaches its denser, critical state.
Thus the massless field is {\em not} part of the critical solution {\it
per se}.  Black hole formation (always with a finite black hole ADM mass
in our study) can take place at times long after the massless pulse has
left the neighborhood of the boson star.  } \label{fig:anim}
\end{figure}

Figure \ref{fig:anim} shows a series of snapshots from a typical
simulation in which the parameter $p$ ($p \equiv A$),
is slightly below the critical
value $p^{\star}$, for a boson star on the stable branch with a mass of $M
= 0.59 M_{Pl}^2/m$ (where $M_{Pl}$ is the Planck mass).   The shell of
massless field, a member of initial data Family $\rm I$, implodes through 
the boson star and explodes back out
from the origin, and the gravitational interaction between the fields
forces the boson star into a new state, a ``critical solution.''  
We see from this animation, and from Figure 3, that
dispersal from the critical state does not mean that the boson star
returns to its original stable configuration, but rather that the star
becomes strongly disrupted and ``explodes.'' That is to say, if we were
to follow the evolution
beyond $t=475$, the massive field would continue to
spread toward spatial infinity.  At some late time, after a large amount
of scalar radiation has been emitted, the end state would probably be a stable
boson star with very low mass. 

\vbox{
\begin{table}
\caption{Families of initial data.  For both families, the initial data, 
$\phi(0,r) = \phi_1(0,r)+i\phi_2(0,r)$,
for the massive complex field 
is given by a boson star, obtained by solving  (\ref{Ham})--(\ref{KG})
 numerically
according to the {\it ansatz} (\ref{BSansatz}) (with the parameter $\omega$
 found via
``shooting'').
The initial real massless field profile, $\phi_3(0,r)$, is given in closed form
 by the ``gaussian'' and
``kink'' initial data.  For each family, we also choose $\partial_t \phi_3(0,r)
$ such
that the pulse is initially in-going, {\i.e.} $\Pi_3(0,r) = \Phi_3(0,r) +
 \phi_3(0,r)/r$.
}
\vspace{0.2cm}
\label{tab:families}
\centerline{ $\begin{array}{c|ccc|ccc}
\hline
\hline
{\rm Family} & \multicolumn{3}{c|}{\rm Complex\ Field\ \phi_1+{\it i}\phi_2}
             & \multicolumn{3}{c}{\rm Real\ Field\ \phi_3} \\
  &  {\rm Name} & {\rm Parameters} & {\rm Profile} 
  &  {\rm Name} & {\rm Parameters} & {\rm Profile}  \\
\tableline
\rm I & {\rm Boson\ Star} & \phi_0(0) & \phi_0(r)  & {\rm Gaussian} &
 A,r_0,\Delta & 
   A \exp\left( - \left({\displaystyle {r - r_0\over \Delta}}\right)^2\right)
 \\
\rm I\,I & {\rm Boson\ Star} & \phi_0(0) & \phi_0(r)  & {\rm Kink} &
 A,r_0,\Delta & 
   {\displaystyle{A\over 2}} \left(1+\tanh\left( {\displaystyle{r - r_0\over
 \Delta}} \right)\right)  \\
\hline
\hline
\end{array}
$}
\end{table}
}

\vspace{-0.4cm}

\begin{figure}
\centerline{ \psfig{figure=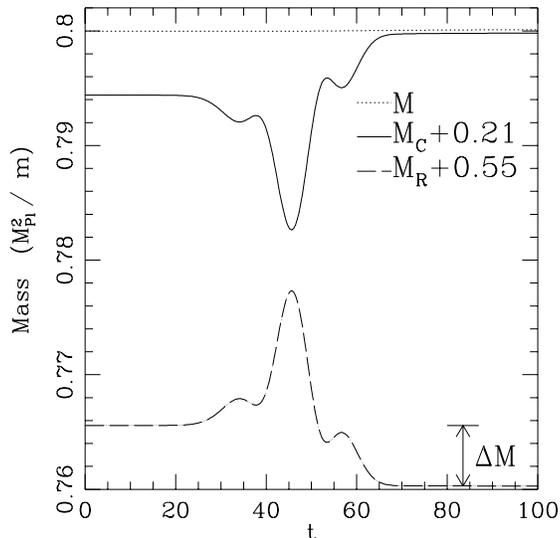,height=7.5cm,width=7.5cm} }
\caption{ 
{Exchange of energy between the real and complex scalar fields.} 
For this simulation, initial data from Family $\rm I$ was used, with 
$\phi_0(0)=0.04\times\sqrt{4\pi}$, $r_0=40$ and $\Delta=8$.
The solid line shows the mass of the complex
field, shifted upward on the graph by $0.21 M_{Pl}^2/m$.  The
long-dashed line shows the mass of the real field, shifted upward by
$0.55 M_{Pl}^2/m$.  
The mass $\Delta M$ exchanged from the massless field to the massive field
in this simulation is nearly 0.0053, or about 2.5\% of the mass of the real 
field (9\% of the boson star mass).   
The amount (and percentage) of mass transfer goes to zero as we consider
boson star initial data approaching the transition to 
instability (see, {\it e.g.} Figure 7).  
The dotted line near the top of the graph shows the
total mass enclosed within $r=100$.  
Throughout the simulation, both the total mass $M = M_C+M_R$ 
and the particle number $N$ (of the complex field) are 
conserved to within a few hundredths of a percent. 
}
\label{fig:massexchange} 
\end{figure}

\begin{figure}
\epsfxsize = 8cm
\epsfysize = 8cm                                            
\centerline{\epsffile{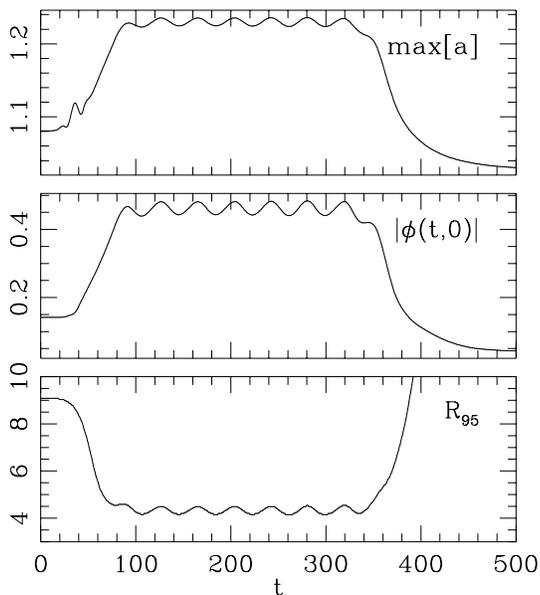}}
\caption{
Quantities describing a near-critical solution.   Here we show timelike
slices through the data shown in Figure \ref{fig:anim}, an evolution
that ends in dispersal.  Top: Maximum value of the metric function $a$ 
(for each spacelike hypersurface parameterized by $t$).
The local maximum at $t\simeq 40$ is due to the presence of the pulse
of massless field.  Middle: Central value $|\phi(t,0)|$ of the massive
field.  Bottom: Radius $R_{95}$ which contains 95\% of the mass-energy
in the complex field.  Again, we see evidence that after the remaining
in critical regime for a while, the star can ``explode,'' leaving a
diffuse remnant with low mass.  } 
\label{fig:crit3_vs_t} 
\end{figure}

\begin{figure}
\epsfxsize = 7.5cm
\epsfysize = 7.5cm
\centerline{\epsffile{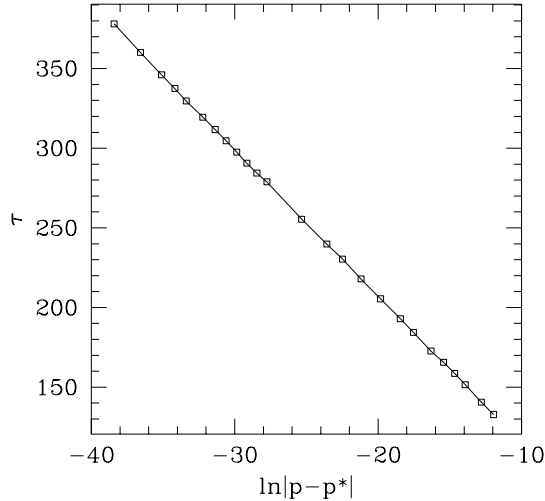}}
\caption{ 
Lifetime $\tau$ of a typical set of near-critical solutions {\it vs.}
$\ln |p-p^\star|$.  We use initial data from Family $\rm I$.
The lifetime of the critical solution
obeys a simple scaling relation.  Using super-critical solutions, we
measure $\tau$ to be the time from $t=0$ until black hole formation occurs. The
relationship shown in the graph can be described by $\tau = -\gamma \ln
|p-p^\star|$, where for the data shown in this graph, $\gamma \simeq 9.2$
The value of $\gamma$
can be related to the imaginary part of the Lyapunov exponent $\sigma$ of the 
unstable mode ($\sim e^{i\sigma t}$) by $\Im(\sigma) = 1/\gamma \simeq 0.109.  
$
This value is the same as that obtained from a linear perturbation analysis
of the specific boson star to which we believe this configuration
is asymptoting (See Section V).
We note that in the limit $p\rightarrow p^\star$, the mass of the black hole
 formed is
finite 
(and close to the mass of the progenitive unstable boson star)
, {\it i.e.} the system exhibits Type I critical behavior.
}
\label{fig:lifetime}
\end{figure}

The gravitational interaction between the two fields
results in an exchange of energy from the
massless field to the massive field, as shown in Figure
\ref{fig:massexchange}.  
Figure \ref{fig:crit3_vs_t} shows some
timelike slices through the simulation data, giving a plot of the
maximum value of $a$, the value of $|\phi|$ at the origin, and the
radius $R_{95}$ as functions of time.  These are given to help
elucidate the point that the critical solution oscillates about some
local equilibrium, before dispersing or forming a black hole.
The lifetime of the critical solution increases monotonically as 
$p\rightarrow p^{\star}$.  Figure \ref{fig:lifetime} shows that the scaling law
 expected
for Type I transitions is exhibited by these solutions.  

\vspace{-0.5cm}
\begin{figure}
\epsfxsize = 7.5cm
\epsfysize = 7.5cm
\centerline{
	\hbox{\psfig{figure=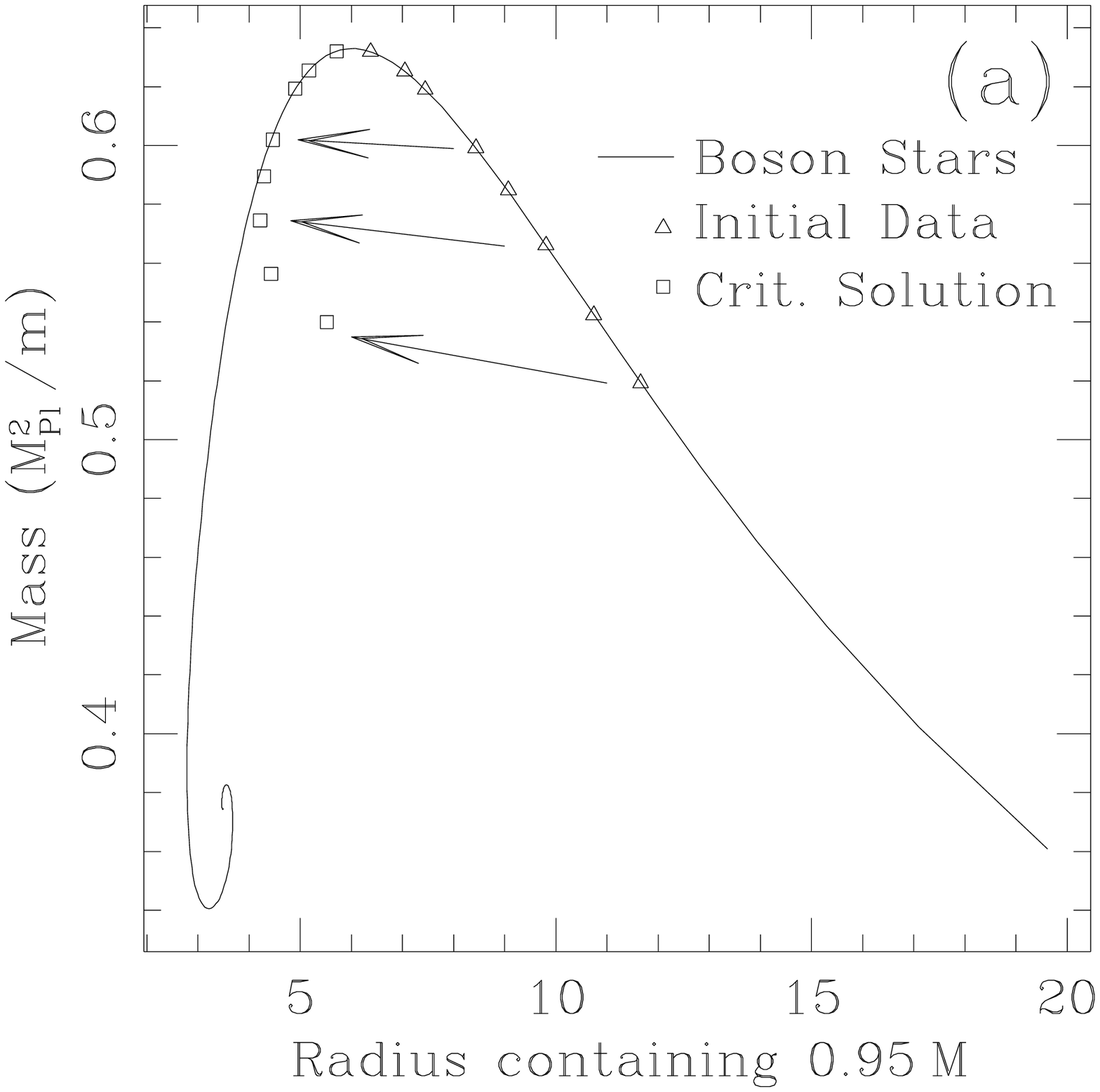,height=7.5cm,width=7.5cm}}
	\hbox{\psfig{figure=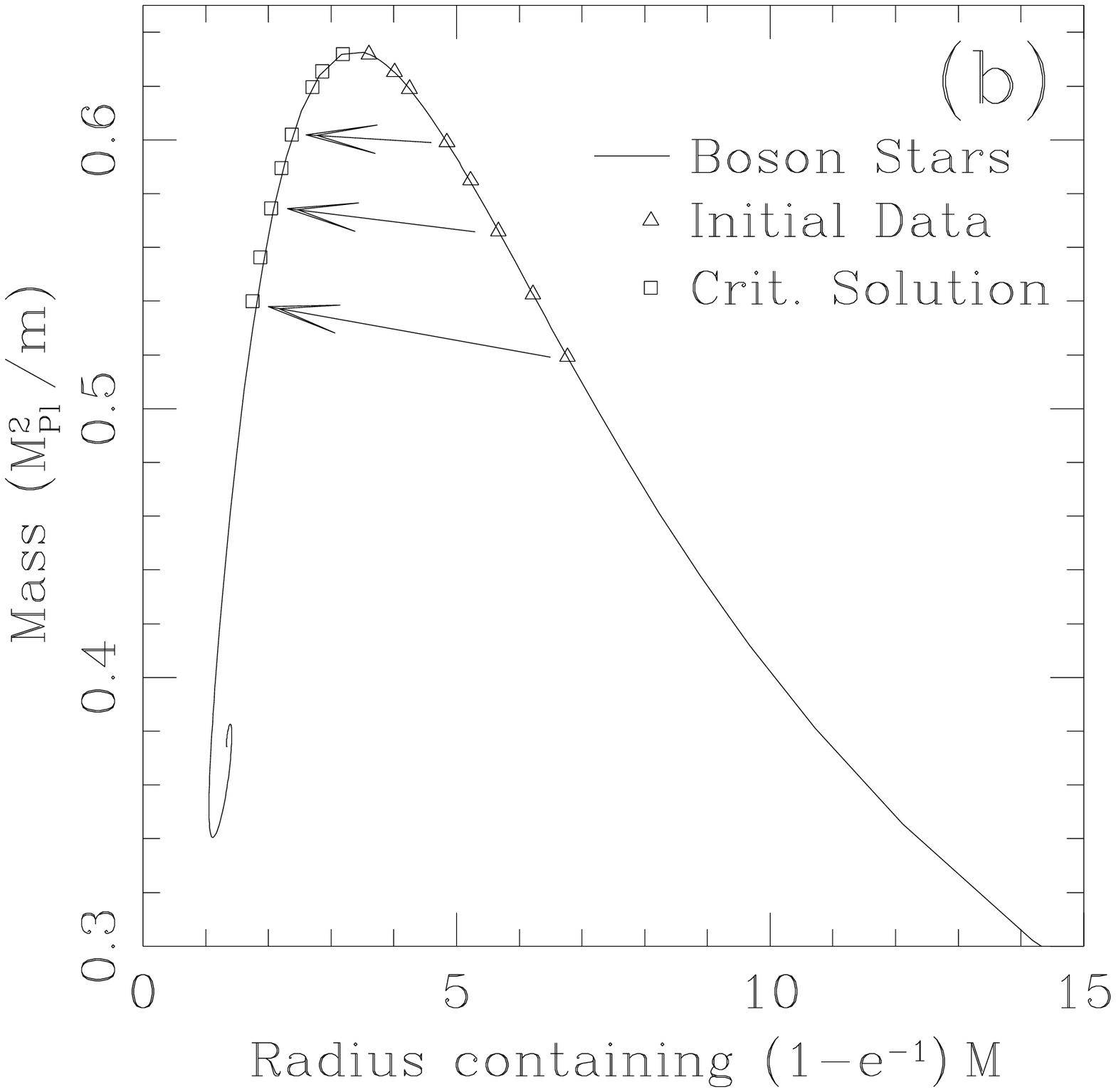,height=7.5cm,width=7.5cm}}
}

\caption{ 
Mass {\it vs.} radius for equilibrium configurations of boson stars
(solid line), initial data for the complex field (triangles), and 
critical solutions (squares).  
Arrows are given to help match initial data with the
resulting critical solutions.  Points on the solid line to the left of
the maximum mass $M_{\rm max}\simeq 0.633 M_{Pl}^2/m$ correspond to unstable
boson stars, whereas those to the right of the maximum correspond to
stable stars.  If one takes time averages of properties such as mass,
central density $|\phi(t,0)|$ and radius $R_{95}$ during the critical
regime, one finds values which match the profile of a boson star on
the unstable branch.  The squares show the time average of each
critical solution during the oscillatory phase.
Graph (a) shows mass $M$ versus $R_{95}$ the radius containing 95\%
of $M$, whereas graph (b) shows $M$ versus the radius containing
$(1-e^{-1})\, M$.  The agreement between the critical solutions and
boson stars shown in graph (a) deteriorates with decreasing mass,
however the comparison shown in graph (b), which neglects the
``tail'' of the critical solutions and boson stars, shows much better
agreement for all masses.  (We show the tail region in Figure
 \ref{fig:showhump}.)
In this simulation the
massive field radiates only a small amount due to the perturbation by
the massless field, and so the stable boson star is essentially driven
to ``pop" across the stability curve by the impinging massless pulse.
}
\label{fig:mvr95} 
\end{figure}

\begin{figure}
\epsfxsize = 7.5cm
\epsfysize = 7.5cm
\centerline{\epsffile{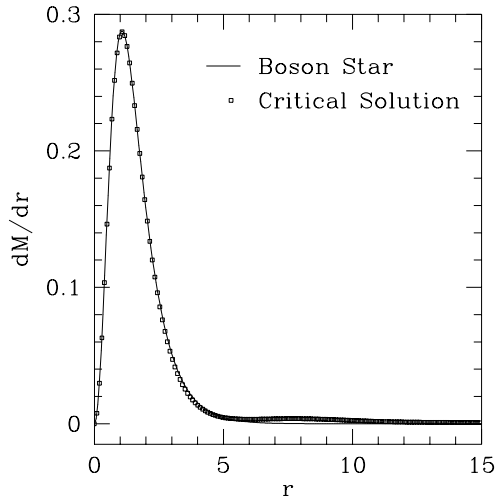}}
\caption{
Comparison of highly unstable (low-mass) critical solution and boson star.  
Squares show a critical solution resulting from a boson star 
having $\phi_0(0) = 0.26\times \sqrt{4\pi}.$  
(The data has been reduced for graphing
purposes; the actual spatial resolution in the simulation is four times
finer than that shown in the figure.) The solid line shows a ``best
fit'' (unstable) boson star we constructed by finding the time average
of $|\phi(t,0)|$ in the critical solution and using this as the value
for $\phi_0(0)$ in the ODE integration 
routine which solves 
for the equilibrium (boson star) solutions.  We see that there is a small halo 
near the outer edge of the solution ($r=8$).
The halo has the same relative magnitude when viewed in
terms of the particle number distribution $\partial N/\partial r$.
We discuss the halo phenomena further in  Section V.
}
\label{fig:showhump}
\end{figure}

Figure \ref{fig:mvr95} shows the mass {\it vs.} radius for some
critical solutions along with the equilibrium curve for boson stars.
We notice that there are great similarities, at least for relatively high 
mass configurations,
between
the critical solutions and unstable boson stars in the ground state.
(We do not perform studies involving boson stars with much lower masses, 
because of the dynamic range required for the spatial resolution of the finite
difference code.  Also, for a given numerical error tolerance, such low-mass 
critical solutions have much shorter lifetimes than larger-mass solutions 
and thus it is more difficult to obtain average properties of such short-lived
critical solutions.)
When we include nearly all of the complex-scalar mass in our 
comparisons, as shown
in Figure \ref{fig:mvr95}(a), we see that the 
time-averaged properties of the critical solutions with lower masses,
{\it i.e.} those further from the transition to instability,
deviate from the curve of equilibrium configurations, and that the deviation
increases as mass decreases.   
When we consider only the bulk of the boson star, however, we see
very good agreement between the dynamically generated critical solutions and 
the unstable boson
stars, computed from the static {\em ansatz}, as shown in 
Figure \ref{fig:mvr95}(b).
The comparison between low-mass critical solutions and boson stars, shown 
in Figure \ref{fig:mvr95},
can be further illuminated by looking at a profile of the mass distribution as
shown in Figure \ref{fig:showhump}.

We see that there is a small halo
near the outer edge of the solution ($r=8$), and it is this which
throws off our measurement of $R_{95}$ used for Figure
\ref{fig:mvr95}.  
Despite the effect this has on the measurement of the
radius $R_{95}$ of the star,  we can still obtain a good fit of a boson
star to the interior of the critical solution in the low-mass regime.
We provide further discussion of these
halos in Section V.
in the critical solutions with higher total mass.  

It is also worth noting that the critical solution best corresponds to
a boson star in the ``ground state,'' {\it i.e.}, without any nodes in the
distribution of the fields $\phi_1$ or $\phi_2$.  Boson stars in 
excited states ({\it i.e., having} nodes in $\phi_1$ and $\phi_2$) have 
mass distributions which differ significantly
from the critical solutions we obtain \cite{BSS}.

We wish to explain these simulation results in terms the quasi-normal
modes of boson stars.  Previous work in critical phenomena
\cite{Mattcrit} - \cite{Brady}, \cite{Carsten} leads us to surmise
that there is a single unstable mode present in the system which is
excited when the boson star moves into the critical regime.  The
oscillatory behavior during the critical regime may be explainable in
terms of the superposition of a stable oscillatory mode with the
unstable mode.  In the next section, we attempt to confirm these
hypotheses by means of perturbation theory.

\section{Boson Star Stability Study via Linear Perturbation Theory}

We follow the work of Gleiser and Watkins \cite{GW}.  
For the perturbation calculations, we find it helpful to 
define the following metric functions:
\begin{eqnarray}
 e^{\nu(t,r)} &\equiv& \alpha^2  \nonumber \\
 e^{\lambda(t,r)} &\equiv& a^2. \nonumber
\end{eqnarray}

and to rewrite the complex field $\phi(t,r)$ as
\begin{equation} \phi(t,r) = [\psi_1(t,r) + i\psi_2(t,r)]e^{-i\omega t},
 \label{phidecomp_pt} \end{equation}
where $\psi_1$ and $\psi_2$ are real.  (Note that this is a different 
decomposition of the field $\phi$ than (\ref{phidecomp}), the one used 
in the previous sections.)

In these variables, the equilibrium quantities are
\begin{eqnarray}
	\lambda(t,r) &=& \lambda_0(r)  \\
	\nu(t,r) &=& \nu_0(r)  \\
	\psi_1(t,r) &=& \phi_0(r)  \\
	\psi_2(t,r) &=& 0.
\end{eqnarray}

For the perturbation, we expand about the equilibrium quantities by
first introducing four perturbation
fields---$\delta\lambda(t,r)$,
$\delta\nu(t,r)$,
$\delta\psi_1(t,r)$
and
$\delta\psi_2(t,r)$---and 
then setting:
\begin{eqnarray}
	\lambda(t,r) &=& \lambda_0(r)  + \delta\lambda(t,r)\\
	\nu(t,r) &=& \nu_0(r)+ \delta\nu(t,r)  \\
	\psi_1(t,r) &=& \phi_0(r) ( 1 + \delta\psi_1(t,r) ) \\
	\psi_2(t,r) &=& \phi_0(r) \delta\psi_2(t,r).
\end{eqnarray}

These expressions are substituted into the relevant evolution and constraint
equations (details in Appendix C), after which the resulting system can be
 reduced
to the following system of two coupled second-order ordinary differential
 equations 
for $\delta\phi_1$ and $\delta\lambda$:

\begin{eqnarray}
\delta\psi_1'' = &-&\left( {2\over r} + {\nu_0'-\lambda_0' \over 2} \right)
 \delta\psi_1'
-{\delta\lambda' \over  r\phi_0^2} + e^{\lambda_0-\nu_0}\delta\ddot{\psi}_1 
 \nonumber \\
&-&\left[ {\phi_0'\over \phi_0}\left( {\nu_0'-\lambda_0' \over 2} + {1\over r}
 \right)
		  + \left({\phi_0'\over \phi_0}\right)^2 +
		  {1- r\lambda_0' \over  r^2 \phi_0^2} +
		  e^{\lambda_0-\nu_0}\omega^2 - e^{\lambda_0} \right]\delta\lambda
  \nonumber \\
&+& 2 e^{\lambda_0} \left[
	 1 + e^{-\nu_0}\omega^2 + e^{-\lambda_0}\left({\phi_0'\over \phi_0}\right)^2
	 +  r \phi_0\phi_0'
\right] \delta\psi_1  \label{d2psidr2}\\
	\delta\lambda'' = &-&{3\over 2}(\nu_0' - \lambda_0') \delta\lambda'
	+\left[ 4 \phi_0'^2 + \lambda_0'' + {2\over r^2} - {(\nu_0' - \lambda_0
')^2\over 2}
			- {2\nu_0' + \lambda_0' \over r} \right] \delta\lambda  
  + e^{\lambda_0-\nu_0}\delta\ddot{\lambda} \nonumber \\
 &-& 4 (2\phi_0\phi_0' - r  e^{\lambda_0}\phi_0^2) \delta\psi_1'  \nonumber \\
 &-& 4\left[ 2\phi_0'^2 - 
		r e^{\lambda_0}\phi_0^2 \left( 2 {\phi_0'\over \phi_0} + {2\nu_0' + \lambda_0
' \over 2}
  \right) \right] \delta\psi_1 .
\label{d2ldr2}
\end{eqnarray}

To perform the stability analysis (normal-mode analysis), we assume a 
harmonic time dependence, {\it i.e.},
\begin{eqnarray}
 \delta\psi_1(t,r) = \delta\psi_1(r)\,e^{i\sigma t} \nonumber \\
 \delta\lambda(t,r) = \delta\lambda(r)\,e^{i\sigma t}.\nonumber
\end{eqnarray}
Note that (\ref{d2psidr2}) and (\ref{d2ldr2}) contain only second 
derivatives with 
respect to time, and because there are good reasons to 
assume $\sigma^2$ is purely real \cite{Jetzer,GW}, we
only need to determine whether $\sigma^2$ is positive or negative to
determine stability or instability, respectively.

\begin{figure}
\epsfxsize = 7.5cm
\epsfysize = 7.5cm
\centerline{\epsffile{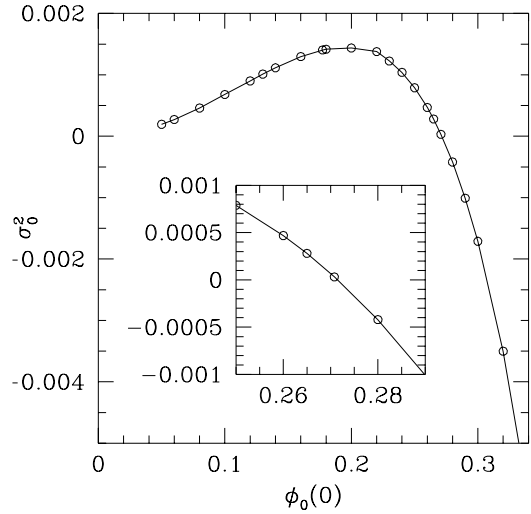}}
\caption{{ Mode frequencies of boson stars: fundamental mode}  
This plot shows a graph of 
$\sigma_0^2$, the squared frequency of the fundamental mode, versus 
the value of $\phi_0$ at the origin.  Note that, as the inset shows, 
$\sigma_0^2$ crosses zero when $\phi_0(0) \simeq 0.27$, which corresponds 
to a boson star with the maximum possible mass.
(The circles show actual values obtained, and the solid line simply
connects these points.)
}
\label{fig:sigsq_vs_phi0}
\end{figure}

Using the method described in Appendix C, we find the distribution 
for the squared frequency $\sigma_0^2$ of the fundamental mode,
with respect to $\phi_0$, which is shown in Figure \ref{fig:sigsq_vs_phi0}.

\begin{figure}
\epsfxsize = 7.5cm
\epsfysize = 7.5cm
\centerline{\epsffile{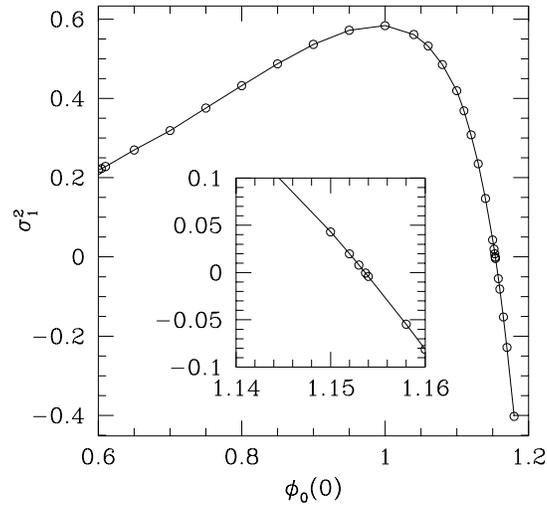}}
\caption{{ Mode frequencies of boson stars: first harmonic mode}  This plot
 shows a graph
of  $\sigma_1^2$, the squared frequency of the first harmonic mode,
versus the value of $\phi_0$ at the origin.  Note that, as the inset
shows, $\sigma_1^2$ crosses zero when $\phi_0(0) \simeq 1.15$, which
corresponds to the first local minimum on the unstable branch of the
mass {\it vs.} radius curve (see Figure 4).  
(The circles show actual values obtained, and the solid line simply
connects these points.)
}
\label{fig:firstharm_vs_phi0}
\end{figure}

Superposed with the fundamental mode, we may have other modes at
higher frequencies.
Figure \ref{fig:firstharm_vs_phi0} shows the relation between
first harmonic frequencies and $\phi_0(0)$.

\vspace{-0.5cm}

\vspace{-0.8cm}

\section{Comparison of Perturbation Analysis and Simulation Data}
We wish to compare the results of our perturbation theory calculation
with the oscillations of stable boson stars.  Two differences exist
between the conventions used in the perturbation theory calculation and
those used in the boson star simulation data.  The first difference is
in the choice of the time coordinate.  In the perturbation theory code,
we choose a lapse of unity at the origin, whereas in the simulations we
set the lapse to unity at spatial infinity.  Thus we have the following
mapping from the perturbation theory calculations to the simulations:

$$ \sigma^2 \Bigl.\Bigr\vert_{\rm Perturbative} \rightarrow 
{\sigma^2 \over \alpha^2} \Bigl.\Bigr\vert_{\rm Simulation}$$

The other significant difference is in the way the complex field $\phi(t,r)$ is
decomposed into constituent real fields.  Thus we cannot directly compare
$\phi_1$ and $\psi_1$, for example.  We can, however, compare the
modulus $|\phi|$ of the field.  For the simulation data, the
perturbation in $|\phi|$ can be taken directly from $(\phi_1^2 +
\phi_2^2)^{1/2}$.  For the data obtained from perturbation theory, the
perturbation in $|\phi|$ will be, to first order,
$\phi_0\delta\psi_1$.

Before proceeding to the comparisons {\em per se}, we wish to point out that
 determining
the unstable mode via numerical simulation of the full {\it nonlinear}
system was very easy to do in comparison to the {\it linear}
perturbation theory calculations.

\vspace{-0.25cm}
\subsection{Unstable modes}
\vspace{-0.25cm}
To measure the unstable mode, we again perform a series of simulations in
which we allow a gaussian pulse from an addition real, massless
Klein-Gordon field to impinge on a stable boson star.  

By tuning the amplitude of this pulse (holding constant the width of the pulse
 and
its initial distance from the boson star), 
we can generate a family of
slightly different near-critical solutions depending on the amplitude of
the initial gaussian pulse, and can tune down the initial magnitude of
the unstable mode.  By subtracting these slightly different near-critical
solutions, we obtain a direct measurement of the unstable mode.

\begin{figure}
\centerline{
	\hbox{\psfig{figure=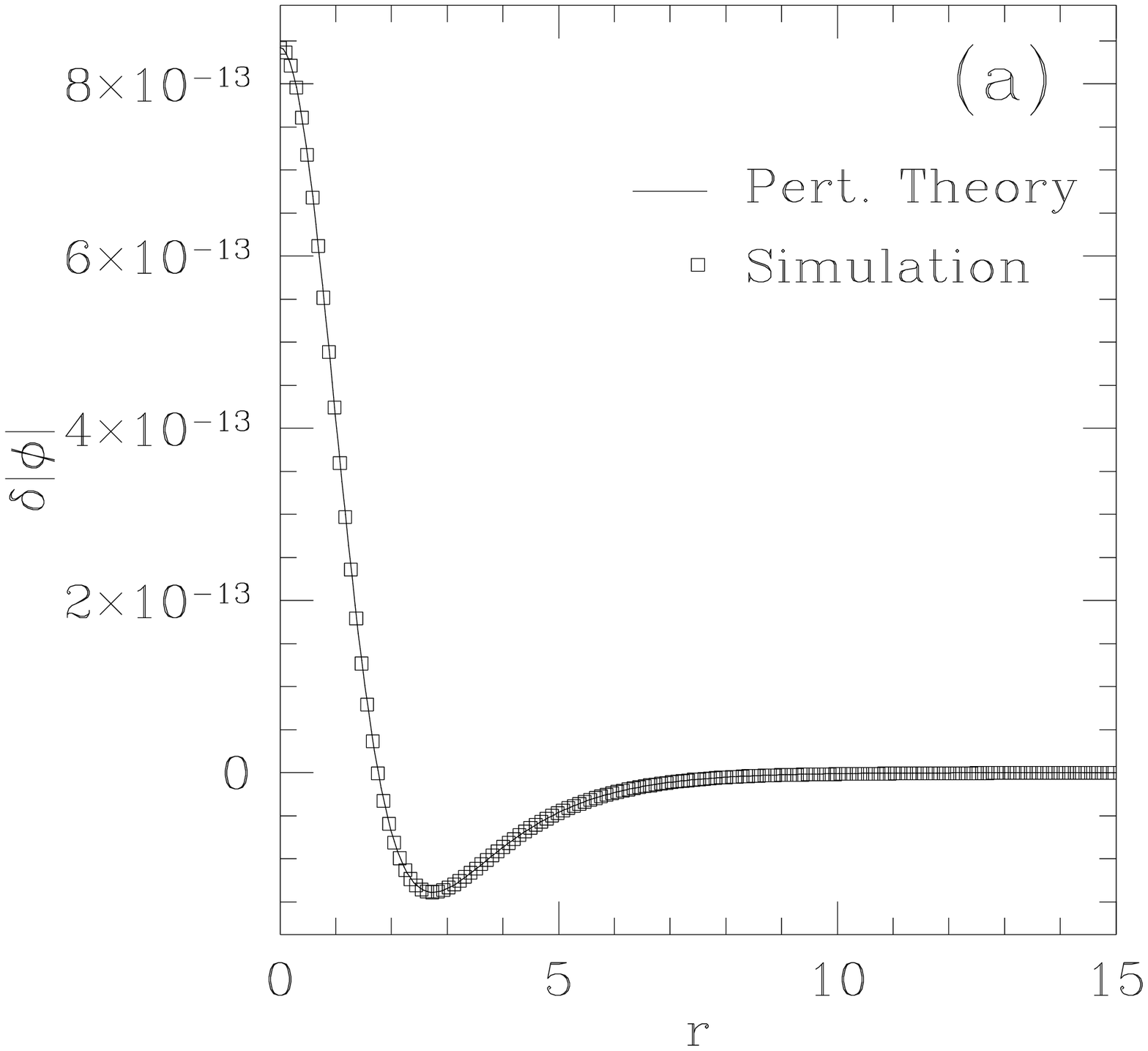,height=7.5cm,width=7.5cm}} 
	\hbox{\psfig{figure=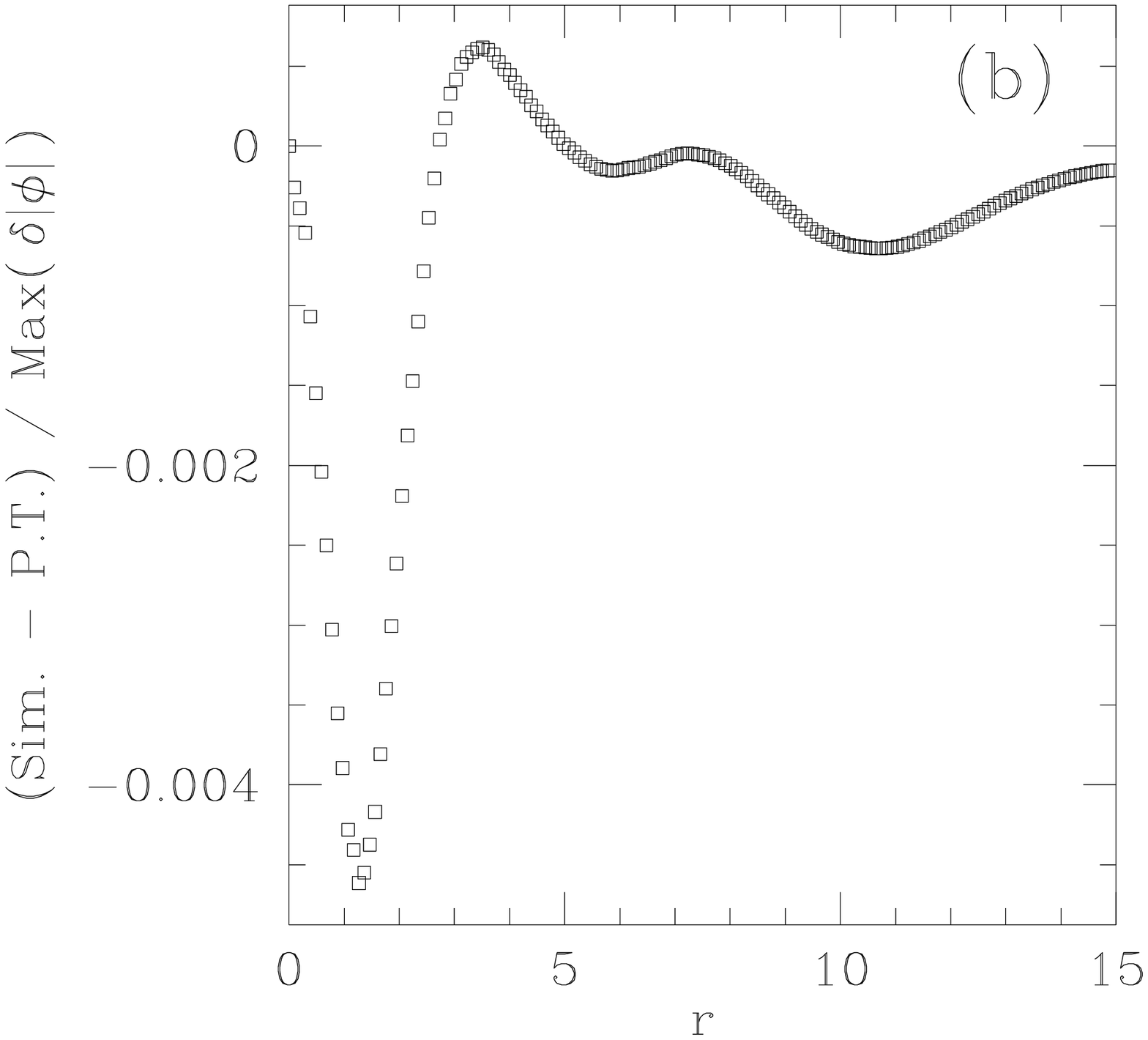,height=7.5cm,width=7.5cm}} 
}
\caption{{ Fundamental mode of unstable boson star.}  
(a) The solid line shows $\phi_0\delta\psi_1$ from the perturbation
theory calculations.  The squares shows the difference between $|\phi|$
for two simulations for which the critical parameter $p$ differs by
$10^{-14}$.  (The data has been reduced for graphing purposes; the
actual spatial resolution in the simulation is four times finer than
what is shown in the figure.) Differences between the simulation data
and perturbation theory results are below $1.1\times10^{-15}$. If a
line were drawn connecting the squares, it would be indistinguishable,
to the eye, from the perturbation theory line.  Thus we provide a
second graph (b) showing the difference of these results, where the
scale is relative to the maximum value of $\delta|\phi|$.  
}
\label{fig:unstab_phi} 
\end{figure}

\begin{figure}
\centerline{
	\hbox{\psfig{figure=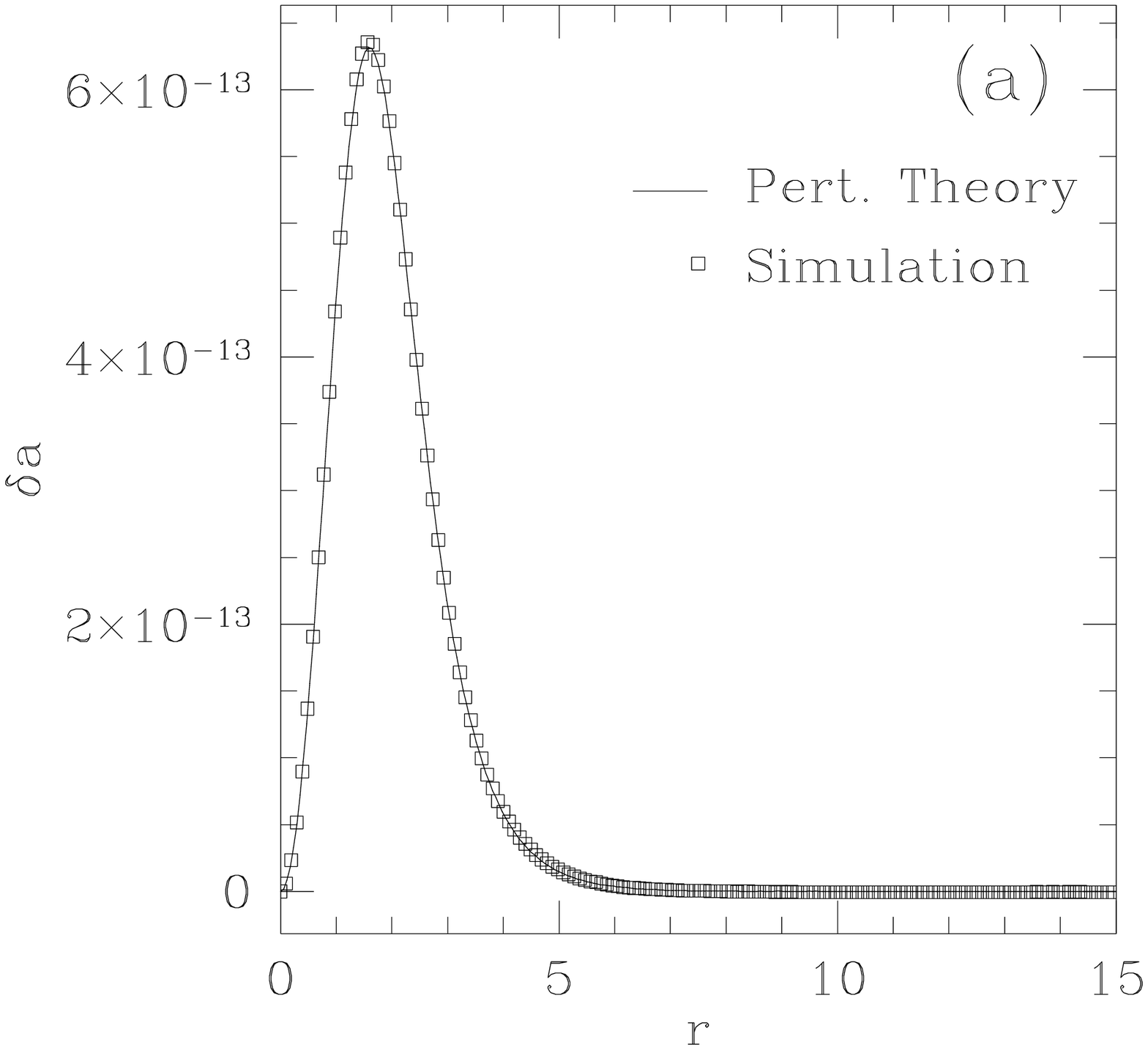,height=7.5cm,width=7.5cm}} 
	\hbox{\psfig{figure=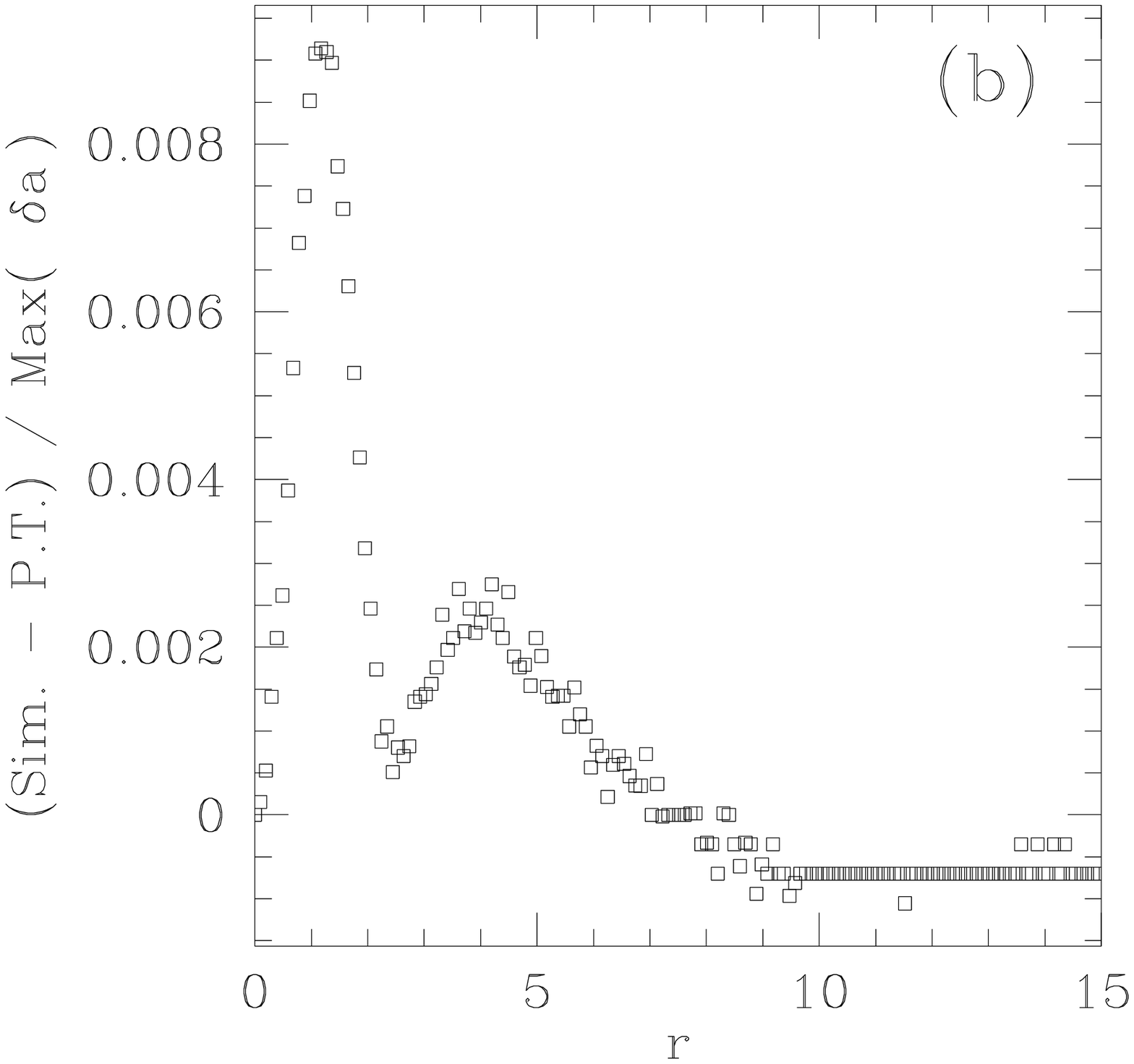,height=7.5cm,width=7.5cm}} 
}
\caption{{ Fundamental mode of unstable boson star.}  
(a) The solid line shows the perturbation to the metric function $a$,
as found from the perturbation theory calculations.  The squares shows
the difference between the metric function $a$ for two simulations for
which the critical parameter $p$ differs by $10^{-14}$. (In the
simulations, the spatial resolution was four times that shown in the
figure.)
(b) A plot of the difference between the mode obtained from the simulation and
the mode obtained via perturbation theory, where the scale is relative to 
the maximum value of $\delta a$.
}
\label{fig:unstab_a}
\end{figure}

\begin{figure}
\epsfxsize = 7.5cm
\epsfysize = 7.5cm
\centerline{\epsffile{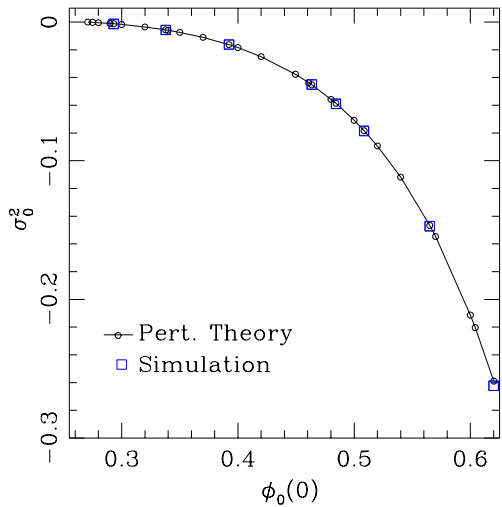}}
\caption{{Comparison of squared frequencies/Lyapunov exponents for 
unstable modes.}  
The circles show a subset of the perturbation
theory data as displayed in Figure \ref{fig:sigsq_vs_phi0}.
The squares show the measurements obtained from our simulations.
(The solid line simply connects the circles.)
We note that the agreement between the two
sets is good even for the more unstable, low-mass solutions.  
We also point out that the measurements of our simulations were 
performed along $r=0$, {\it i.e.,}, in the interior of the halo 
found in the low-mass solutions, which seems to strengthen the remarks 
at the end of section III, namely that, aside
from the halo at the exterior of the critical solution, the critical
solutions (of all masses) seem to correspond to unstable boson stars.
}
\label{fig:freq_comp}
\end{figure}

Considering a specific example, 
we start with a stable boson star which has an initial field value at the
origin of $\phi_0(0)=0.04\times\sqrt{4\pi}$.  By driving it with a
gaussian pulse tuned to machine precision, we can cause this stable
star to become a critical solution which persists for very long times,
oscillating about a local equilibrium.  The average value of
$|\phi(t,0)|$ is $\langle|\phi_(t,0)|\rangle \simeq 0.463$.  We measure
the unstable mode by subtracting data of a run which contained a
gaussian pulse with an amplitude that differed by $10^{-14}$ from that
of the pulse tuned to machine precision.  We can then measure the
growth factor of the unstable mode by taking the $L_2$ norm of this
difference at various times, taking the logarithm, and fitting a
straight line to it.  From this, we obtain $\sigma \simeq 0.109\,i$, or
$\sigma^2 \simeq -0.0118$.  Because of the differences in time
coordinate between the simulations and perturbation theory
calculations, we need to compute $\sigma^2/\alpha^2$ in order to
compare with the perturbation calculations.  We find the average value
of $1/\alpha(t,0)^2$ for the times listed above to be $\langle
1/\alpha(t,0)^2\rangle \simeq 3.80$, and thus we find ${\sigma^2 /
\alpha^2} \simeq -0.0450.$ We choose to compare these perturbation
theory results with data from a time in the simulation for which the
difference in field values (for the two evolutions tuned slightly
differently) is $\Delta|\phi(t,0)|\simeq 8.4\times 10^{-13}$.  We use
this value in the perturbation theory solver and arrive at $\sigma^2
\simeq -0.045$, in good agreement with the value from the simulation.
In Figures \ref{fig:unstab_phi} and \ref{fig:unstab_a}, we compare the
graphs of the solutions for the unstable mode.
In Figure \ref{fig:freq_comp} we show a comparison of the squared
freqency values obtained from the linear perturbative analysis and
those as measured in our simulations.  

\vspace{-0.8cm}

\vspace{-0.4cm}

\subsection{Oscillatory modes}
We can also look at the oscillatory mode during the critical regime.
We study the behavior of such a mode using the same
technique we used to examine the fundamental mode of
the unstable boson star: we subtract the data at one instant of time from
the data at all other instants.  
Again, as a specific example,
we use the same initial boson star as
that used in the previous section.
During the critical portion of the evolution,
we notice an oscillation period of about $T \simeq 38.4$, and thus we
obtain $\sigma = 2\pi /T\simeq 0.0261.$ During this period, the average
value of $1/\alpha^2(t,0)$ is about $3.80$, and thus we find $\sigma^2
/ \alpha^2 \simeq 0.102.$ We take data from a moment in the 
middle of the oscillation period, and subtract it from the data at other times.
We can then compare the perturbation theory results with simulation
data at a local  peak of the oscillation.  For the local peak we chose
at time $t=t_p$, the difference in modulus of the field was
$\Delta|\phi_(t_p,0)| \simeq 0.0197$.  Inserting this value into the
perturbation theory code, we find $\sigma^2 \simeq 0.102$ for this
configuration.  Thus we again find excellent agreement between 
the squared oscillation frequencies computed
in perturbation theory and via simulation.

\begin{figure}
\centerline{
	\hbox{\psfig{figure=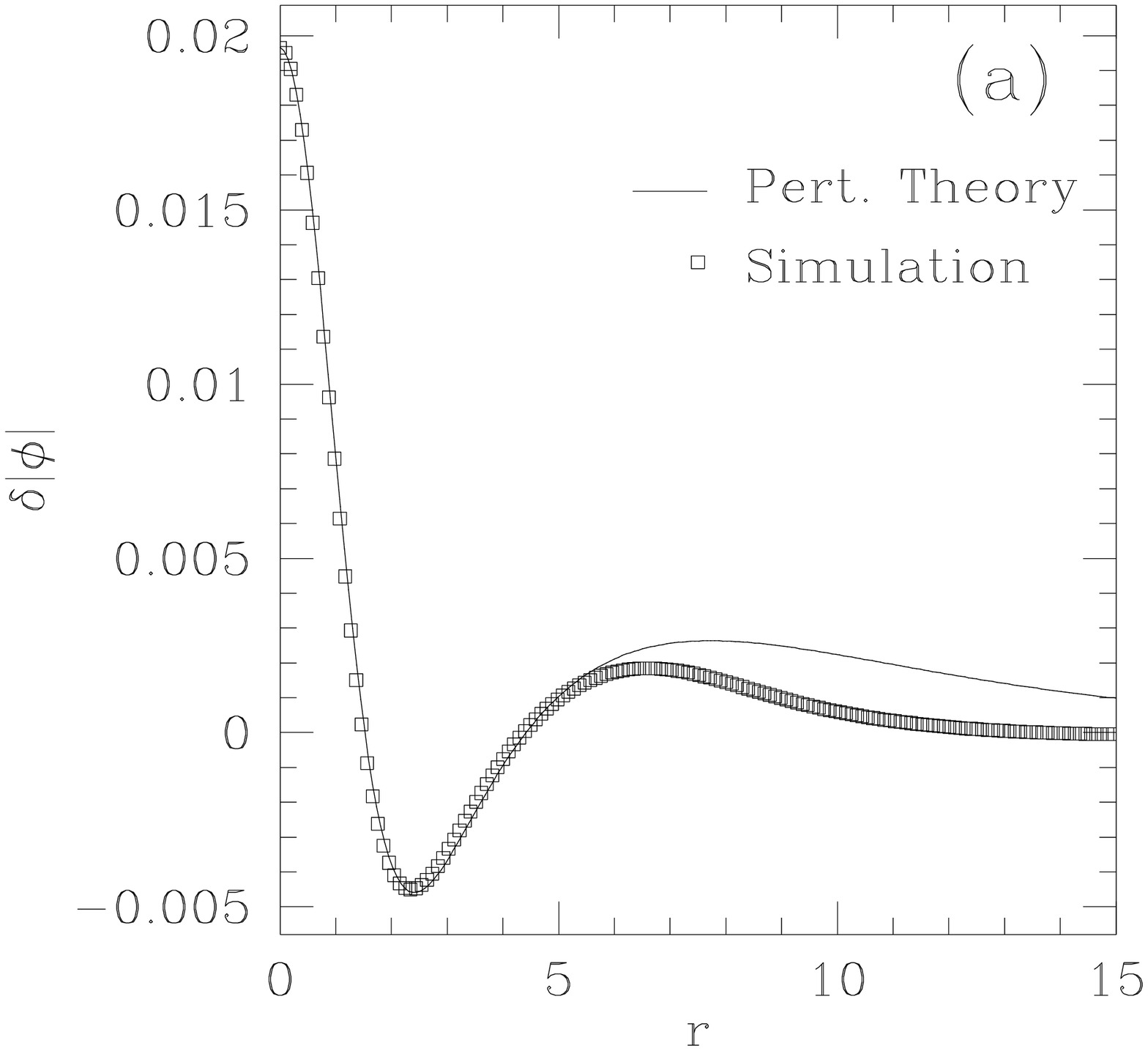,height=7.6cm,width=7.6cm}} 
	\hbox{\psfig{figure=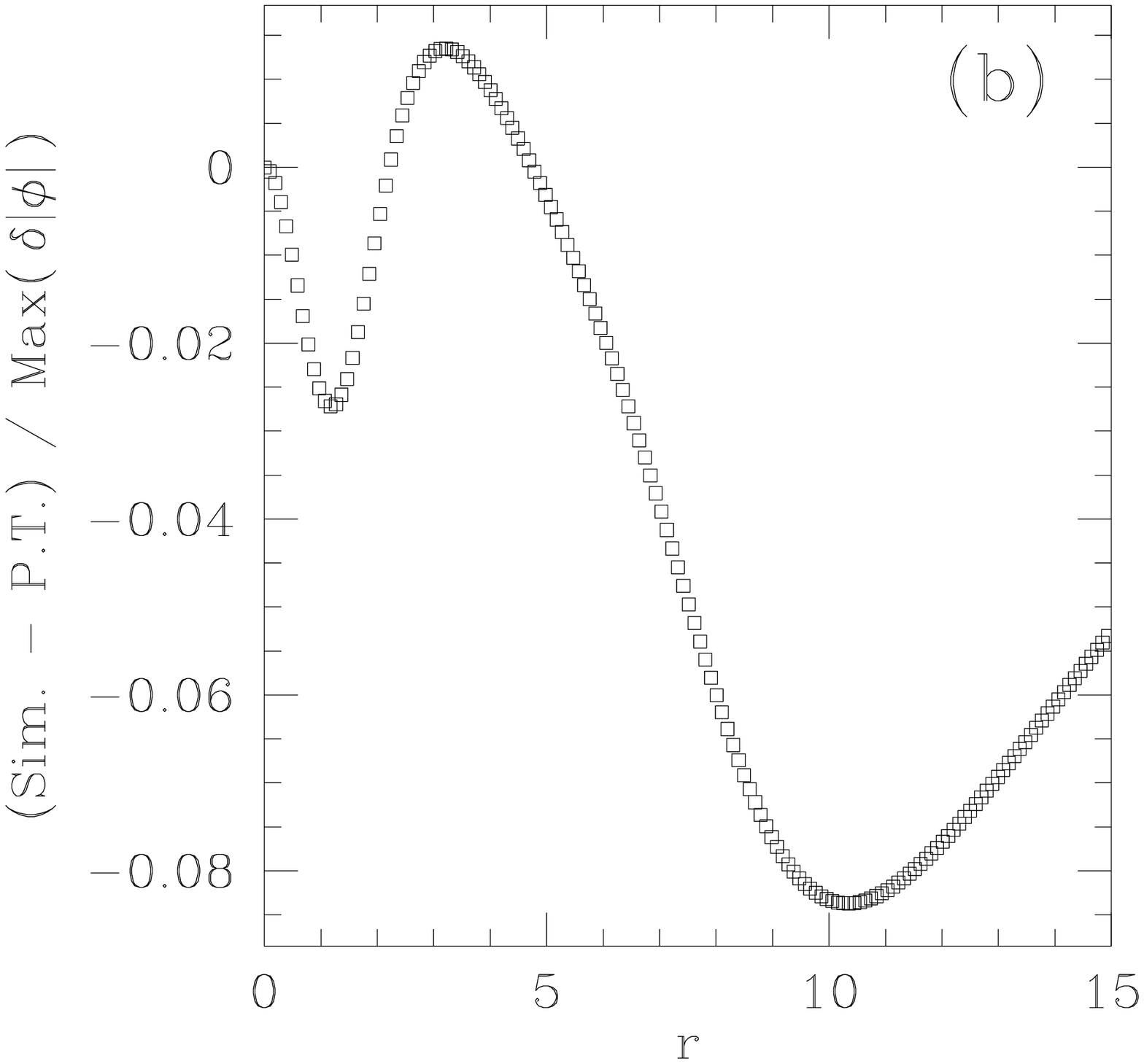,height=7.6cm,width=7.6cm}} 
}
\caption{{ First harmonic of an unstable boson star.}  
(a) The solid line shows $\phi_0\delta\psi_1$ from the perturbation
theory calculations.
To obtain the squares, we took the simulation data and 
subtracted the Klein-Gordon field at one instant of time from the data at 
another instant.  (The data in the simulations had a spatial resolution four
times finer than what is shown in the figure.)
(b) The squares show the difference between mode obtained
from simulation and the mode obtained via perturbation theory,
scaled relative to the maximum value of $\delta |\phi|$.
As we describe in the text, the lack of agreement
beyond $r\simeq 6$ may be an artifact of simplistic data analysis.
The next figure shows that the metric quantities, which depend directly
on the matter distribution (and thus on $|\phi|$), show a favorable
comparison between the simulations and perturbation theory.
}
\label{fig:firstharm_phi}
\end{figure}

\begin{figure}
\centerline{
	\hbox{\psfig{figure=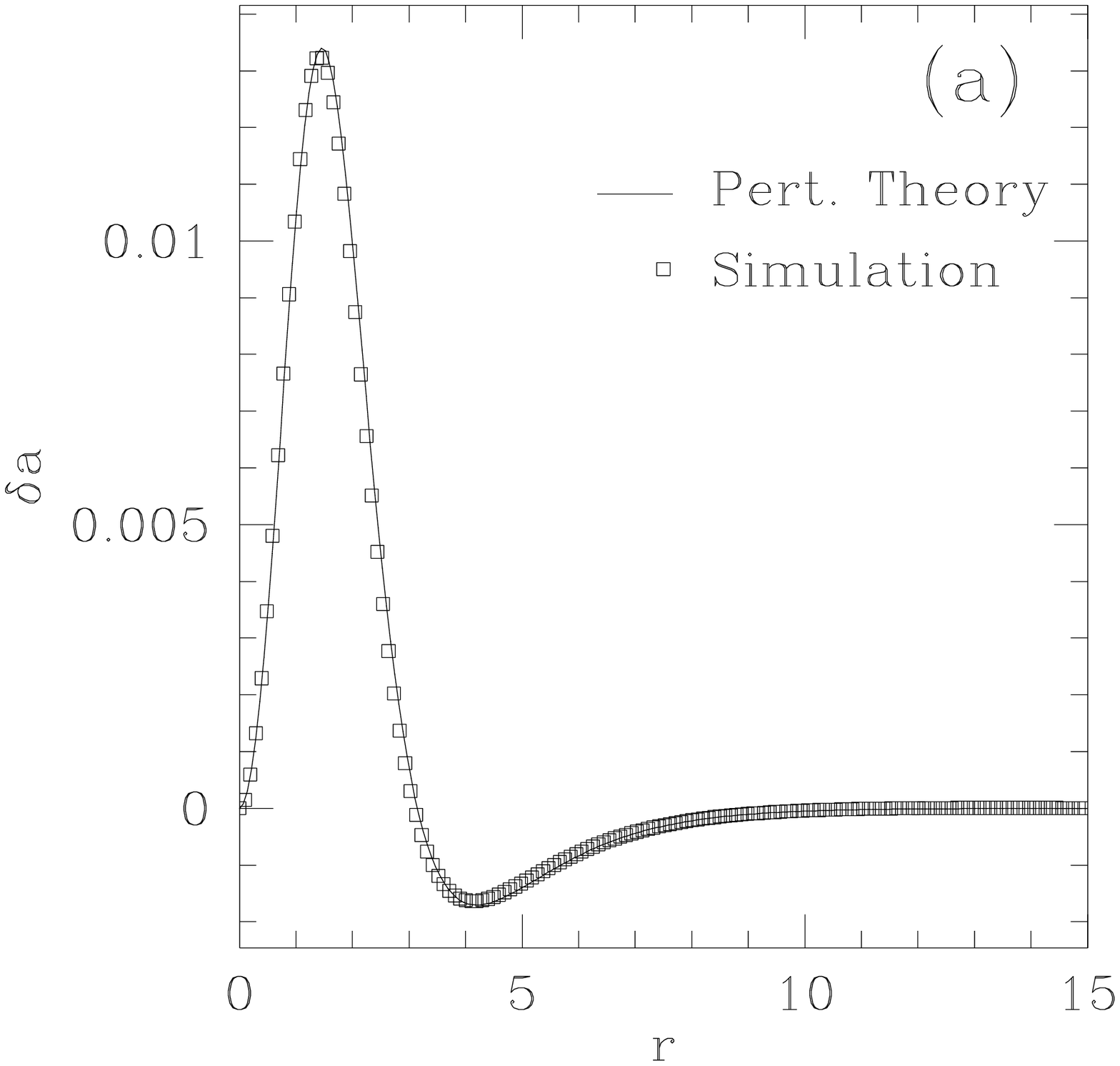,height=7.6cm,width=7.6cm}} 
	\hbox{\psfig{figure=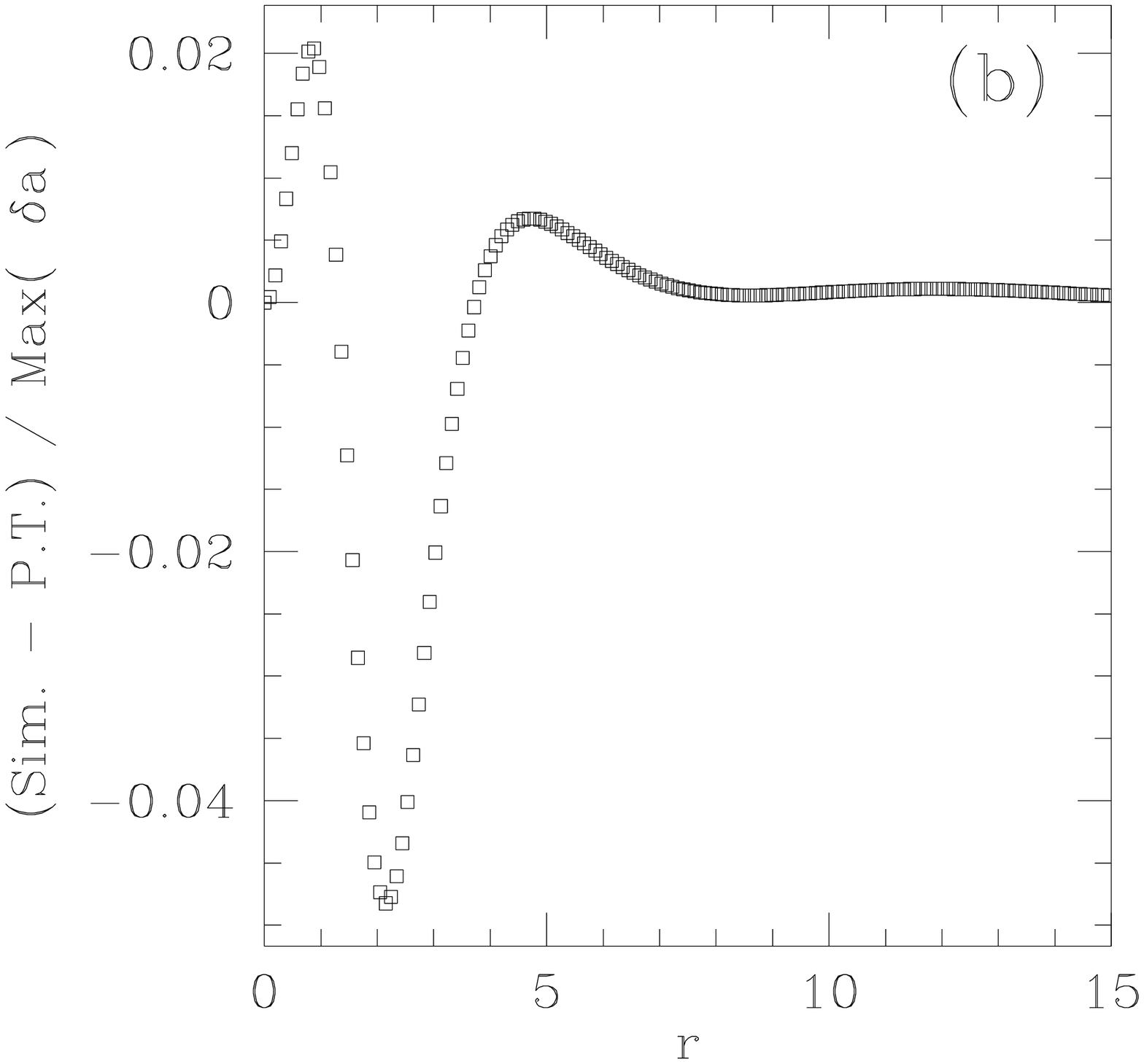,height=7.6cm,width=7.6cm}} 
}
\caption{{ First harmonic of an unstable boson star.}  
(a) The solid line shows the perturbation to $a$  as found
from perturbative calculations.  
To plot the squares, we took the simulation data 
and subtracted the metric function $a$ 
at one instant of time from the data at another instant.
(The spatial resolution in the simulation was four times finer than what
is shown in the figure.)
(b) The squares show the difference between the simulation data and the results
of linear perturbation theory, scaled relative to the maximum value of
$\delta a$.
The close fit between these results indicates that the oscillations observed in
the critical solutions correspond to stable oscillatory modes in an unstable
 boson star.
}
\label{fig:firstharm_a}
\end{figure}

In Figures \ref{fig:firstharm_phi} and \ref{fig:firstharm_a}, 
we compare the functions obtained from the perturbation theory 
calculation with those from the simulation.
We note that the agreement for the metric functions is very good for
all radii, but the agreement in the fields begins to decline beyond
$r=5$.  Why do the graphs of $|\phi|$ not agree well for the first
harmonic?  This could be a consequence of our simplistic method of
extracting this mode.  While our method of simply subtracting different
frames has worked well for our test cases of oscillations of stable
boson stars, the first harmonic of the unstable star has a higher
frequency and thus our graph could be subject to sampling effects.  A
better method would be to perform a Fourier transform in time for each
grid point, and construct the higher harmonics in the field
accordingly.  
There may be a simple resolution to the discrepancy in
the graphs of $|\phi|$, 
(\ref{phidecomp}) and 
agreement in the graphs of the metric, our analysis does seem to
indicate that the oscillations observed for this data in fact
correspond to the first harmonic quasinormal mode of a boson star,
however the analysis of the matter field needs further attention.

Finally, we must remark that we have been unable, using the fundamental
and first harmonic modes of an unstable boson star, to construct a
solution possessing a halo similar to that shown in Figure
\ref{fig:showhump}.  We do {\em not} expect higher modes to be of any use
here, because the halo is observed to oscillate with the same (single)
frequency as the rest of the star.  Since, as we described at the end
of section III, the halo seems to be radiated away over time, we
might not expect it to be described by the quasinormal modes (which conserve
particle number) we have constructed.   

\section{Halos}
We have strong evidence that that the critical solutions correspond 
to unstable boson stars, but the principal point of disagreement is
existence of a ``halo'' of massive field which resides in the ``tail''
of the solution.  It is our contention that
this halo is not part of the true critical solution, but rather, is an
artifact of the collision with the massless field.    

In particular, the halo
seems to be a remnant of the original (stable) boson star which is not induced
to collapse with the rest of the star to form the true critical solution.
We find that such a halo is observable in nearly all but
the most massive (least unstable) critical solutions we have considered,
and that its size tends to increase as less massive (more unstable) 
solutions are generated.  The fact that the halo thus {\em decreases} as we 
approach the turning point only makes sense---a stable boson star very 
close to the turning point needs very little in the way of a perturbation
from the massless field to be "popped" over to the unstable branch, and 
the final, unstable configuration, will, of course, be very close to 
the initial state.

Additionally, we note that in all cases we have exmained, the field 
comprising the halo 
oscillates with nearly the same
(single) frequency as the rest of the solution. This indicates that the halo
is not explainable in terms of additional higher-frequency modes.

\begin{figure}
\centerline{
        \hbox{\psfig{figure=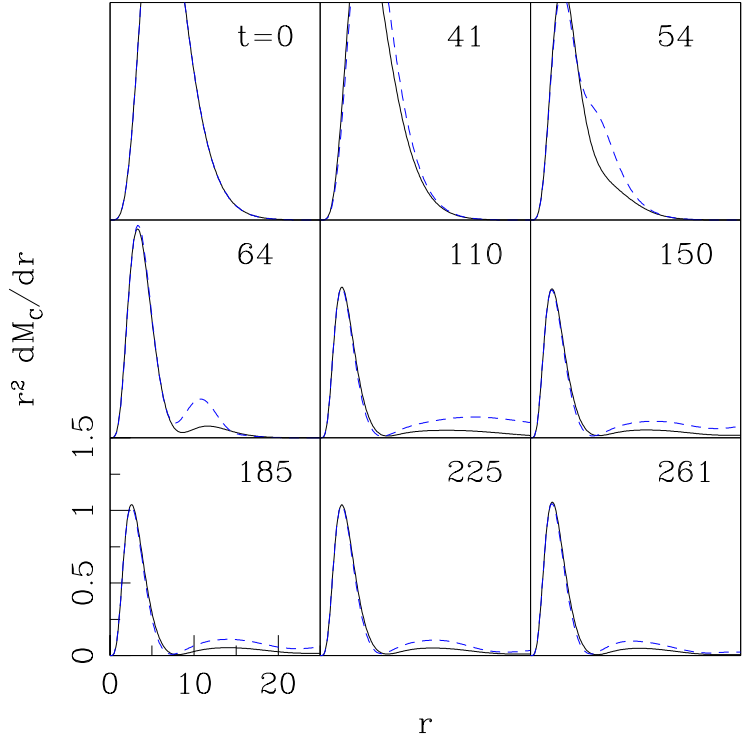,height=10cm,width=10cm}}
}
\caption{
Evolution of $r^2 dM_C/dr$ for for two different sets of initial data.
Both sets contain the same initial boson star, but the massless field $\phi_3$
for one set is given by a ``gaussian" of Family I (solid line) with $r_0=30$,
 and $\Delta=8$
whereas for the other set $\phi_3$
is given by a ``kink" of Family II (dashed line) with $r_0=35$ and $\Delta=3$.
The variable $A$ is varied (independently for each family) as the parameter $p$
to obtain the critical solution.
(Note that after $t\simeq 60$, the massless field has completely left the
 domain shown 
in the figure.)
We have multiplied $dM_C/dr$ by $r^2$ to highlight the dynamics of the halo;
 thus the main
body of the solution appears to decrease in size as it moves to lower values of
 $r$.
The kink data produces a larger and much more dynamical halo, but interior
to the halo, the two critical solutions match closely --- and also match the
 profile of
an unstable boson star.
Thus, the portion of the solution which is ``universal'' corresponds to an
 unstable boson
star.
}
\label{fig:haloanim}
\end{figure}

\begin{figure}
\centerline{
        \hbox{\psfig{figure=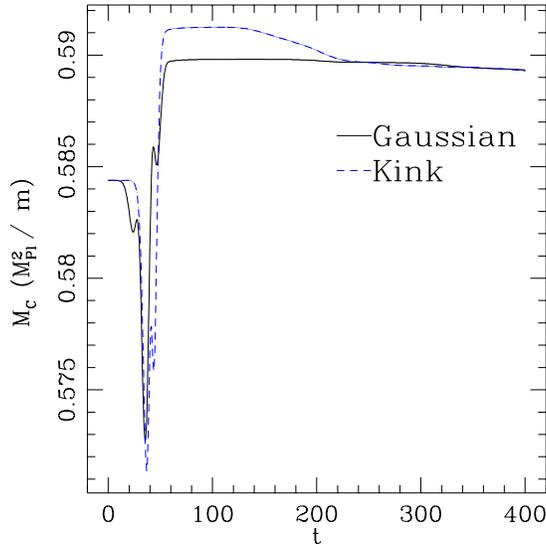,height=7.6cm,width=7.6cm}}
}
\caption{ $M_C$ {\it vs.} time for the two evolutions shown in Figure
 \ref{fig:haloanim}.
Mass transfer from the real to the complex field 
occurs from $t\simeq 30$ to $t\simeq 60$, {\it i.e.} while the supports of the
 fields overlap.
There is more mass transferred using the kink data, and yet the mass falls off
 rapidly.
The mass of the kink data acquires a value very close to the mass of the
 gaussian 
data, which is itself decreasing slowly with time.  We see that, beyond 
$t\simeq 250$,
the difference in mass between the two solutions is very small compared with
 the
amount of mass transferred from the real field.
}
\label{fig:Mc_vs_t.ps}
\end{figure}

As one might expect, 
the properties of the halo are not universal, {\it i.e.} they are quite
dependent on the type of initial data used.  In contrast, the critical
solution interior to the halo is largely independent of the form of the
initial data. 
To demonstrate this, we use two families of initial data, given by
a ``gaussian" of Family $\rm I$ in Table \ref{tab:families} and  a ``kink" 
of Family $\rm I\,I$.  A series of snapshots from one such pair of evolutions
is shown in Figure \ref{fig:haloanim}.
We find different amounts of mass transferred from the massless to the massive
field for the kink and gaussian data, as shown in Figure \ref{fig:Mc_vs_t.ps},
yet the central values of the field oscillate about nearly
the same value at nearly the same frequency.
Both calculations start with identical 
boson stars with $|\phi(0,0)|=0.04\times\sqrt{4\pi}$.
In the critical regimes, 
this  becomes $\langle |\phi(t,0)|\rangle=0.130\times\sqrt{4\pi}$ for the
 solution 
obtained from
the gaussian data, and $\langle|\phi(t,0)|\rangle=0.135\times\sqrt{4\pi}$ for
 the kink data.
As already noted, the oscillation periods are also quite similar, differing by
 about $3\%$, and the massess 
interior to the halo are also quite comparable. In particular, it seems quite 
remarkable that the differences in mass interior to the
halo for the two families are much smaller than the mass transferred from 
the real field in either case.

\begin{figure}
\epsfxsize = 7.5cm
\epsfysize = 7.5cm
\centerline{\epsffile{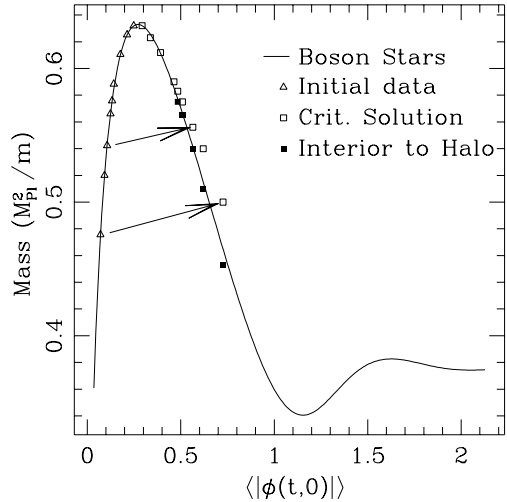}}
\caption{
Mass {\it vs.} $\langle |\phi(t,0)| \rangle$, the time average of the
central value of the field for equilibrium configurations of boson
stars (solid line), initial data (triangles) and critical solutions
(open and filled squares).  Arrows are given to help match initial data
with the corresponding critical solution.  Points on the solid line to
the left of the maximum mass $M_{\rm max}\simeq 0.633 M_{Pl}^2/m$
correspond to stable boson stars, whereas those to the right of the
maximum correspond to unstable stars.  The data is the same as that
used for Figure \ref{fig:mvr95}, with data from one further evolution
added at the bottom of the mass range.  The open squares show the time
average of the mass and $|\phi(t,0)|$ of some critical solutions, and
the filled squares show the same quantities evaluated between $r=0$ and the
inner edge of the halo, defined to be the point where
$\partial|\phi|/\partial r = 0$ for finite $r$.  
  The mass of the critical solution is in general
greater than the mass of the initial data, however the mass inside the
halo of the critical solution is less than the mass of the initial
data.
}
\label{fig:mass_vs_cd}
\end{figure}

If we consider the inner edge of the halo to be where
$\partial|\phi|/\partial r = 0$ at some finite radius 
({\it e.g.,} $r\simeq 5$ in Figure \ref{fig:showhump}), and look at the data
between $r=0$ and the inner edge of the halo, we find good agreement
between this data and the profile of a boson star.  This can be seen
in both Figures \ref{fig:showhump} and \ref{fig:mass_vs_cd}.

We suspect that the halo is radiated over time (via scalar radiation, or 
``gravitational cooling'' \cite{SandSGC}) 
for all critical solutions.  We find, however, that the time scale for
 radiation of the
halo is comparable to the time scale for dispersal or black hole formation for
 each 
(nearly) critical solution we
consider.  Thus, while we see trends which indicate that the halo is indeed
 radiating,
we are not able to demonstrate this conclusively for a variety of scenarios.
With higher numerical precision, one might be able to more finely
tune out the unstable mode, allowing more time to observe the behavior of
the halo before dispersal or black hole formation occur.

\section{Conclusions}
We have shown that it is possible to induce gravitational collapse and,
in particular, Type I critical phenomena in spherically-symmetric boson
stars in the ground state, by means of ``perturbations'' resulting from
gravitational interaction with an in-going pulse from a massless real scalar
field.   Through this interaction, energy is transferred from the real to
the complex field, and complex field is ``driven'' and ``squeezed'' to
form a critical solution.
The massless field is not directly involved in the critical behavior
observed in the complex massive field;  the critical solution itself
appears to correspond to a boson star, which, at any 
finite distance from criticality in parameter space,  exhibits a superposition
 of
stable and unstable modes.

Specifically, for initial data consisting of a boson star with nearly the
 maximum
possible mass of $M_{\rm max}\simeq 0.633 M_{pl}^2/m$, the resulting
critical solution oscillates about a state which has all the features
of the corresponding unstable boson star in the ground state, having
the same mass as the initial star.   This result is reminiscent of the
study by Brady {\it et al.} \cite{Brady}, who found that the Type I
critical solutions for a real massive scalar field corresponded to the
oscillating soliton stars of Seidel and Suen \cite{SandS}.  
For boson stars with a mass somewhat less than $M_{\rm max}$, {\it e.g.}, $0.9
M_{\rm max}$  or less, however, we find less than complete agreement
between the critical solution and an unstable boson star of comparable
mass.  This is evidenced by the presence of an additional spherical
shell or ``halo'' of matter in the critical solution, located in what
would be the tail of the corresponding boson star.  
Interior to this halo, we find  that the critical solution compares 
favorably with the
profile of an unstable boson star.  Additionally, we have shown 
that the halo details depend on the specifics of the perturbing 
massless field, and we conjecture that, in the infinite time 
limit, the halo would be radiated away.

In order to extend the comparison between the critical solutions and
boson stars, we have verified and applied the linear perturbation analysis
presented by Gleiser and Watkins \cite{GW}, extending their work by
providing an algorithm to obtain modes with nonzero frequency.  We have used
this algorithm to give quantitative distributions of mode frequency
{\it vs.} central density of the boson star for the first two modes, as
well as to solve for the modes to compare with our simulation results.
We have found that the unstable mode in the critical solutions have the same
growth rate as the unstable mode of boson stars, and that the mode shapes
also compare quite favorably.  We noted that the unstable mode of these
boson stars was determined {\em much more easily} by solving the full
nonlinear set of evolution equations, rather than via linear
perturbation theory.  The oscillations observed in the critical
solution also indicated agreement with first harmonic mode obtained via
perturbation theory, however the oscillatory mode in $|\phi|$ showed
poor agreement at large radii, and awaits more careful analysis.

Future work may include simulations of the critical solutions of low
mass using higher numerical precision to further tune away the initial
amplitude of the unstable mode, thus allowing more time to observe the
the small halo ({\it i.e.}, whether it is in fact being radiated
away).  We would also hope to obtain better agreement between
simulation and perturbation theory for the first harmonic mode of the
field $|\phi|$, perhaps using a more sophisticated method of extracting
modes from the simulation.  
Another direction worthy of note would be to begin the simulation
with a pulse of the complex field (instead of specifically a boson star)
tune the height of the pulse to find the critical solutions via interpolation,
and then compare the resulting critical solutions with 
our results obtained by perturbing boson stars.

Finally,
we find it worthwhile to investigate similar scenarios for neutron
stars.  While there have been studies regarding the explosion of
neutron stars near the minimum mass ({\it e.g.}, \cite{ENS1},
\cite{ENS2}), we would like to see whether neutron stars of
{\it non-minimal mass} can be driven to explode via dispersal from a 
critical solution.   This may take the form of a neutron star approaching
the onset of instability via slow accretion, or by being driven across
the stability graph via violent heating from some other matter source,
in a manner similar to the perturbations of boson stars we have considered
in this paper.

\acknowledgments
This research was supported by NSERC and by NSF PHY9722068
Some computations were carried out on the {\tt vn.physics.ubc.ca}
Beowulf cluster which was funded by the Canadian Foundation
for Innovation.  Other calculations were performed using the Cray T3E
at the Texas Advanced Computing Center.

\appendix
\section{Boson Star Mode Frequencies}

In this Appendix we have tabulated some sample values from the
perturbation theory calculations.   The values and uncertainties
expressed in the table captions were determined by integrating
(\ref{d2psidr2}) and (\ref{d2ldr2}) to various maximum radii, for a
range of error tolerances in the integration routines.  The values and
uncertainties given in the tables were chosen to express the variation
in our results.

\vspace{0.4cm}
\tabcolsep 0.07in
\vbox{
\begin{table}
\caption{Shooting Parameters: Fundamental Mode.
The values of $\phi_0(0)$ are exact.
Other quantities are given within an uncertainty of $\pm 1$
in the last significant digit.
}
\label{tab1}
\begin{tabular}{cccr@{.}l}
$\phi_0(0)$ &  $\omega$ &  $\delta\lambda''(0)/\delta\psi_1(0)$  & 
\multicolumn{2}{c}{$\sigma^2$}\\
\tableline
\tt 6.0E-02 & \tt   1.0417E+00   & \tt    1.68E-01   & \tt    0&\tt 28E-03   \\
\tt 1.0E-01 & \tt   1.0727E+00   & \tt    0.29E+00   & \tt    0&\tt 67E-03   \\
\tt 1.4E-01 & \tt   1.1067E+00   & \tt    0.43E+00   & \tt    1&\tt 11E-03   \\
\tt 1.8E-01 & \tt   1.1440E+00   & \tt    0.59E+00   & \tt    1&\tt 41E-03   \\
\tt 2.2E-01 & \tt   1.1849E+00   & \tt    0.77E+00   & \tt    1&\tt 31E-03   \\
\tt 2.6E-01 & \tt   1.2299E+00   & \tt    0.98E+00   & \tt    0&\tt 45E-03   \\
\tt 2.7E-01 & \tt   1.2419E+00   & \tt    1.04E+00   & \tt    0&\tt 05E-03   \\
\tt 2.8E-01 & \tt   1.2542E+00   & \tt    1.10E+00   & \tt   -0&\tt 43E-03   \\
\tt 3.0E-01 & \tt   1.2796E+00   & \tt    1.24E+00   & \tt   -1&\tt 71E-03   \\
\tt 4.0E-01 & \tt   1.4281E+00   & \tt    2.08E+00   & \tt   -1&\tt 84E-02   \\
\tt 5.0E-01 & \tt   1.6215E+00   & \tt    3.45E+00   & \tt   -7&\tt 09E-02   \\
\tt 6.0E-01 & \tt   1.8777E+00   & \tt    5.79E+00   & \tt   -2&\tt 11E-01  \\ 
\end{tabular}
\end{table}
}

\vbox{
\tt
\begin{table}
\caption{Shooting Parameters: First Harmonic Mode.
The values of $\phi_0(0)$ are exact,
$\omega$ is given within an uncertainty of $\pm 1$
in the last significant digit, and the other quantities
are given within an uncertainty of $\pm 2$
in the last significant digit.
}
\begin{tabular}{cccr@{.}l}
$\phi_0(0)$ &  $\omega$ &  $\delta\lambda''(0)/\delta\psi_1(0)$  & 
\multicolumn{2}{c}{$\sigma^2$}\\
\tableline
\tt 6.00E-01  & \tt  1.8777E+00   & \tt    0.63E+01   & \tt   0& \tt 22E+00 \\
\tt 7.00E-01  & \tt  2.2230E+00   & \tt    1.13E+01   & \tt   0& \tt 32E+00 \\
\tt 8.00E-01  & \tt  2.6963E+00   & \tt    2.09E+01   & \tt   0& \tt 43E+00 \\
\tt 9.00E-01  & \tt  3.3536E+00   & \tt    4.11E+01   & \tt   0& \tt 53E+00 \\
\tt 1.00E+00  & \tt  4.2714E+00   & \tt    0.84E+02   & \tt   0& \tt 54E+00 \\
\tt 1.10E+00  & \tt  5.5471E+00   & \tt    1.77E+02   & \tt   0& \tt 42E+00 \\
\tt 1.12E+00  & \tt  5.8555E+00   & \tt    2.07E+02   & \tt   3& \tt 05E-01 \\
\tt 1.14E+00  & \tt  6.1842E+00   & \tt    2.41E+02   & \tt   1& \tt 46E-01 \\
\tt 1.15E+00  & \tt  6.3566E+00   & \tt    2.59E+02   & \tt   4& \tt 30E-02 \\
\tt 1.16E+00  & \tt  6.5346E+00   & \tt    2.80E+02   & \tt  -8& \tt 11E-02 \\
\tt 1.17E+00  & \tt  6.7184E+00   & \tt    3.02E+02   & \tt  -2& \tt 28E-01 \\
\tt 1.18E+00  & \tt  6.9083E+00   & \tt    3.26E+02   & \tt  -4& \tt 01E-01 \\ 
\end{tabular}
\end{table}
}

\section{Finite Difference Algorithm}
We approximate the continuum field quantities $\lbrace{ \alpha, a,
\Pi_1, \Pi_2, \Pi_3, \Phi_1, \Phi_2, \Phi_3, \phi_1, \phi_2, \phi_3
\rbrace}$ by a set of {\it grid functions}, quantities which are
obtained via the solution of finite difference approximations to the
partial differential equations (\ref{defPhi}), (\ref{Ham}) -
(\ref{PhiPirel}) on a domain which has been discretized into a regular
mesh ({\it i.e.} lattice) with mesh spacing $\Delta r$ in space and
$\Delta t$ in time.  For a grid function $u$, we denote the value of
the grid function in the mesh location $j$ in space and $n$ in time by
$u^{{}^{\scriptstyle n}}_{{}_{\scriptstyle j}}$, {\it e.g,} 
$$ \alpha^{{}^{\scriptstyle n}}_{{}_{\scriptstyle j}} \simeq
 \alpha\left(n\Delta t,(j-1)\Delta r\right),$$ 
where $\alpha\left(n\Delta t,(j-1)\Delta r\right)$ is the corresponding
value for the continuum solution.

The initial data is obtained via ``shooting,'' a standard method of
solving ordinary differential equations, in a way essentially the same
as that found in \cite{RB}.  The numerical method used for evolving the
system of equations is a {\it leapfrog scheme}, which is an explicit
scheme requiring data at two previous time steps, $n$ and $n-1$, to
compute a value at the next time step $n+1$.  Given a discretization of
scale of order $h$ in time and space, the leapfrog scheme is ${\mathcal
O}(h^2)$ accurate.  Throughout the mesh, the ratio 
$\lambda_{\tt CFL} \equiv \Delta t / \Delta r$ 
is kept at a constant value, which must be less than unity due to the
stability requirements of the leapfrog scheme.

To aid in the presentation of the difference equations, we define the
following operators \cite{MattDiss}:

$$ \Delta^{{}^{\scriptstyle t}}_{{}_{\scriptstyle 0}} u^{{}^{\scriptstyle
 n}}_{{}_{\scriptstyle j}} = 
								 { u^{{}^{\scriptstyle n+1}}_{{}_{\scriptstyle j}} -
 u^{{}^{\scriptstyle n-1}}_{{}_{\scriptstyle j}} \over 2 \Delta t}  $$
$$ \Delta^{{}^{\scriptstyle r}}_{{}_{\scriptstyle 0}} u^{{}^{\scriptstyle
 n}}_{{}_{\scriptstyle j}} = 
								  { u^{{}^{\scriptstyle n}}_{{}_{\scriptstyle j+1}} -
 u^{{}^{\scriptstyle n}}_{{}_{\scriptstyle j-1}} \over 2 \Delta r} $$
$$ \Delta^{{}^{\scriptstyle r}}_{{}_{\scriptstyle +}} u^{{}^{\scriptstyle
 n}}_{{}_{\scriptstyle j}} = 
								  { u^{{}^{\scriptstyle n}}_{{}_{\scriptstyle j+1}} -
 u^{{}^{\scriptstyle n}}_{{}_{\scriptstyle j}} \over \Delta r} $$
$$ \Delta^{{}^{\scriptstyle r}}_{{}_{\scriptstyle 3}} u^{{}^{\scriptstyle
 n}}_{{}_{\scriptstyle j}} = 
	3 { u^{{}^{\scriptstyle n}}_{{}_{\scriptstyle j+1}} - u^{{}^{\scriptstyle
 n}}_{{}_{\scriptstyle j-1}} \over (r^{{}^{\scriptstyle }}_{{}_{\scriptstyle j
+1}})^3 - (r^{{}^{\scriptstyle }}_{{}_{\scriptstyle j-1}})^3}. $$
We also define the averaging operator
$$ \mu^{{}^{\scriptstyle r}}_{{}_{\scriptstyle +}} u^{{}^{\scriptstyle
 n}}_{{}_{\scriptstyle j}} =  {1\over 2} \left( u^{{}^{\scriptstyle
 n}}_{{}_{\scriptstyle j+1}} + u^{{}^{\scriptstyle n}}_{{}_{\scriptstyle j}}
 \right), $$
which takes precedence over other algebraic operations, {\it e.g.}
$$ \mu^{{}^{\scriptstyle r}}_{{}_{\scriptstyle +}} \left( f g^2\over h \right) 
= 
 { \mu^{{}^{\scriptstyle r}}_{{}_{\scriptstyle +}} f^{{}^{\scriptstyle
 n}}_{{}_{\scriptstyle j}} \left( \mu^{{}^{\scriptstyle r}}_{{}_{\scriptstyle 
+}}g^{{}^{\scriptstyle n}}_{{}_{\scriptstyle j}} \right)^2 
		  \over \mu^{{}^{\scriptstyle r}}_{{}_{\scriptstyle +}}h^{{}^{\scriptstyle
 n}}_{{}_{\scriptstyle j}} }
\, .$$

The evolution equations, which are applied to each field 
$\lbrace\Phi_i,\Pi_i, i=1,2,3\rbrace$ can then be written as:
\begin{equation}
 \Delta^{{}^{\scriptstyle t}}_{{}_{\scriptstyle 0}} \Phi^{{}^{\scriptstyle
 n}}_{{}_{\scriptstyle j}} = 
				\Delta^{{}^{\scriptstyle r}}_{{}_{\scriptstyle 0}} \left( {\alpha \over a}
 {\Pi} \right) ^{{}^{\scriptstyle n}}_{{}_{\scriptstyle j}}
\end{equation}
\begin{equation}
  \Delta^{{}^{\scriptstyle t}}_{{}_{\scriptstyle 0}} \Pi^{{}^{\scriptstyle
 n}}_{{}_{\scriptstyle j}} = 
			 \Delta^{{}^{\scriptstyle r}}_{{}_{\scriptstyle 3}} \left( {r^2\alpha\over
 a}\Phi \right)^{{}^{\scriptstyle n}}_{{}_{\scriptstyle j}}
			 - 2\left(\alpha a\phi\right)^{{}^{\scriptstyle n}}_{{}_{\scriptstyle j}}
\end{equation}
where the last term in the evolution equation for $\Pi$ is not applied
to the massless field.

Our boundary conditions are as follows: 
First, by regularity at the origin, we have 
$$\Phi^{{}^{\scriptstyle n}}_{{}_{\scriptstyle 1}} = 0$$
for all $n$.
To obtain $\Pi^{{}^{\scriptstyle n+1}}_{{}_{\scriptstyle 1}}$
we employ a ``quadratic fit'' at the advanced time,
\begin{equation}
\Pi^{{}^{\scriptstyle n+1}}_{{}_{\scriptstyle 1}} = { 4 {\Pi}^{{}^{\scriptstyle
 n+1}}_{{}_{\scriptstyle 2}} - {\Pi}^{{}^{\scriptstyle n+1}}_{{}_{\scriptstyle
 3}} 
		 \over 3},
\end{equation}
which is based on the regularity condition,
$\lim_{r\to0} \Pi(t,r) = \Pi_0(t) + r^2 \Pi_2(t) + \, \cdots$.

A significant challenge in the numerical solution of these equations is
the problem of the outer boundary condition for the massive field.
Numerous authors have proposed methods to handle this.  Having tried
various methods including first order expansions of the dispersion
relation \cite{SandS}, sponge filters \cite{IandO}, and operator
splitting \cite{Arbona}, we were unable to obtain a scheme which
produced results superior to the simple Sommerfeld condition one uses
for massless fields \cite{MyWeb}.  Since, however, the Sommerfeld
condition is still inadequate for massive fields, we have chosen to run
our simulations on a grid large enough that the outer boundary is out
of causal contact with the region of interest for the time the
simulation runs.  So, for example, if we are interested in a region
$0\le r\le 50$ and times $0\le t\le 400$, then we place the outer
boundary $r_J \ge 450$.  (While unbounded phase velocities are
a feature of the Klein-Gordon equation, we can argue on physical
grounds as well as see quite clearly in simulations that it is the
group velocity which is the important quantity in the numerical
evolutions, and this is sub-luminal.)   Recent work using a shifted
coordinate system, with a shift vector that is vanishing in some region
near $r=0$ but increases to unity as $r\rightarrow r_J$, shows promise
as a means of handling the challenge of the boundary condition for the
massive field \cite{Ethan}, and this method may be employed in future
work.  Thus the outer boundary condition we employ is \cite{ChopTech}:

\begin{equation}
{\Phi}^{{}^{\scriptstyle n+1}}_{{}_{\scriptstyle J}} = 
 \left( {3\over \Delta t} + {3\over \Delta r} + {2\over r^{{}^{\scriptstyle
 }}_{{}_{\scriptstyle J}}} \right)^{-1}
 \left(
  { 4{\Phi}^{{}^{\scriptstyle n}}_{{}_{\scriptstyle J}} -
 {\Phi}^{{}^{\scriptstyle n-1}}_{{}_{\scriptstyle J}} \over \Delta t}
 + {4{\Phi}^{{}^{\scriptstyle n+1}}_{{}_{\scriptstyle J-1}} -
 {\Phi}^{{}^{\scriptstyle n+1}}_{{}_{\scriptstyle J-2}} \over \Delta r}
 \right)
\end{equation}
and an analagous equation is used for each $\Pi$ variable.

After these evolved variables are obtained at the $n+1$ time step,
we apply a form of numerical dissipation advocated by Kreiss and Oliger
\cite{KO}.  This is applied to both ${\Phi}^{{}^{\scriptstyle n
+1}}_{{}_{\scriptstyle j}}$ and
${\Pi}^{{}^{\scriptstyle n+1}}_{{}_{\scriptstyle j}}$ in the same manner.  So,
 for instance we
set
\begin{equation}
{\Phi}^{{}^{\scriptstyle n+1}}_{{}_{\scriptstyle j}}:= {\Phi}^{{}^{\scriptstyle
 n+1}}_{{}_{\scriptstyle j}} -
  {\epsilon\over 16}\left(
  {\Phi}^{{}^{\scriptstyle n-1}}_{{}_{\scriptstyle j+2}} -
 4{\Phi}^{{}^{\scriptstyle n-1}}_{{}_{\scriptstyle j+1}}
 + 6 {\Phi}^{{}^{\scriptstyle n-1}}_{{}_{\scriptstyle j}}
 - 4 {\Phi}^{{}^{\scriptstyle n-1}}_{{}_{\scriptstyle j-1}} +
 {\Phi}^{{}^{\scriptstyle n-1}}_{{}_{\scriptstyle j-2}}
  \right),
\end{equation}
where $\epsilon$ $(0<\epsilon<1)$ is an adjustable parameter:
typically, we use $\epsilon = 0.5$.

The preceeding equations describe the ``evolution'' aspect of the code.
The other variables are evolved in a ``constrained'' manner, {\it i.e.}
they are obtained on the spacelike hypersurface $n+1$ after the fields
$\Phi^{{}^{\scriptstyle n+1}}_{{}_{\scriptstyle j}}$ and $\Pi^{{}^{\scriptstyle
 n+1}}_{{}_{\scriptstyle j}}$ have been 
calculated.
The field values $\phi^{{}^{\scriptstyle n+1}}_{{}_{\scriptstyle j}}$ are
 obtained by updating the 
value at the outer boundary $j=J$ according to
\begin{equation}
  \Delta^{{}^{\scriptstyle t}}_{{}_{\scriptstyle 0}}\phi^{{}^{\scriptstyle
 n}}_{{}_{\scriptstyle J}} =
	 + \left({\alpha \over a}\,\Pi\right)^{{}^{\scriptstyle n}}_{{}_{\scriptstyle
 j}}
\end{equation}
and then integrating {\it inward} from $j=J$ to $j=1$ along the spatial 
hypersurface at $n+1$: 
\begin{equation}
 \Delta^{{}^{\scriptstyle r}}_{{}_{\scriptstyle +}}\phi^{{}^{\scriptstyle
 }}_{{}_{\scriptstyle j}} = \mu^{{}^{\scriptstyle r}}_{{}_{\scriptstyle +}}
 \Phi^{{}^{\scriptstyle }}_{{}_{\scriptstyle j}}.
\end{equation}

The Hamiltonian constraint (\ref{Ham}) can be solved at each time step
once all the field variables have been computed for the advanced time
step.  We use the variable $A\equiv \ln\,a$ to avoid loss of precision
near the origin in the following finite difference approximation, which is
evaluated at the advanced time step $n+1$:
\begin{equation}
\Delta^{{}^{\scriptstyle r}}_{{}_{\scriptstyle +}} A^{{}^{\scriptstyle
 }}_{{}_{\scriptstyle j}} = 
  \mu^{{}^{\scriptstyle r}}_{{}_{\scriptstyle +}}\left( 
			{1-e^A \over 2\, r}
			+ {r\over 2} \left[ \Pi_1^2 + \Pi_2^2 + \Pi_3^2 +
										\Phi_1^2 + \Phi_2^2 + \Phi_3^2 +
				  e^A \left( \phi_1^2 + \phi_2^2 \right) \right] 
	 \right)^{{}^{\scriptstyle }}_{{}_{\scriptstyle j}} .
\label{eq:DiffHam}
\end{equation}

This equation is solved using a {\it pointwise} Newton iteration, {\it
i.e.} given a value of $A^{{}^{\scriptstyle n+1}}_{{}_{\scriptstyle j}}$ (such
 as $A^{{}^{\scriptstyle n+1}}_{{}_{\scriptstyle 1}}=0$ at
the origin), we find the next value $A^{{}^{\scriptstyle n
+1}}_{{}_{\scriptstyle j+1}}$ outward along the
spatial hypersurface by solving (\ref{eq:DiffHam}) via Newton's
method.

The slicing condition can be solved once the field variables and the metric 
function $a$ have been obtained at the advanced time step, using
the following linear algebraic relation:
\begin{equation}
	 \alpha^{{}^{\scriptstyle n+1}}_{{}_{\scriptstyle j+1}} =
 \alpha^{{}^{\scriptstyle n+1}}_{{}_{\scriptstyle j}} \cdot 
			  { (1/\Delta r) + Z \over (1/\Delta r) - Z}\, ,
\label{eq:DiffSlice}
\end{equation}
where
$$Z \equiv  \mu^{{}^{\scriptstyle r}}_{{}_{\scriptstyle +}} \left( a^2-1 \over
 2 r\right)^{{}^{\scriptstyle }}_{{}_{\scriptstyle j}}
	 + {\Delta^{{}^{\scriptstyle r}}_{{}_{\scriptstyle +}} a^{{}^{\scriptstyle
 }}_{{}_{\scriptstyle j}} \over \mu^{{}^{\scriptstyle r}}_{{}_{\scriptstyle 
+}}a^{{}^{\scriptstyle }}_{{}_{\scriptstyle j}} }
	 - \mu^{{}^{\scriptstyle r}}_{{}_{\scriptstyle +}}\left[
			 ra^2 m^2 \left( \phi_1^2 + \phi_2^2 \right)
	 \right]^{{}^{\scriptstyle }}_{{}_{\scriptstyle j}} 
.$$

\section{Details of Linear Stability Analysis}
Following Gleiser and Watkins \cite{GW}, we write the most general
time-dependent, spherically-symmetric 
metric as
$$ ds^2 = -e^{\nu(t,r)}dt^2 + e^{\lambda(t,r)}dr^2 + r^2d\Omega,$$
and decompose the complex massive field $\phi(t,r)$ via
\begin{equation} \phi(t,r) = [\psi_1(t,r) + i\psi_2(t,r)]e^{-i\omega t},
 \end{equation}
where $\psi_1$ and $\psi_2$ are real.  

In these variables, the Hamiltonian constraint and slicing condition can 
be written as

\begin{equation}
\lambda' = {1-e^\lambda\over r} +  r\left(
	e^{\lambda-\nu}\left[ (\dot{\phi}_1 + \omega\psi_2)^2 
								  + (\dot{\phi}_2 - \omega\psi_1)^2 \right]
	+ \psi_1'^2 + \psi_2'^2 +  e^{\lambda}(\psi_1^2+\psi_2^2) \right)
\label{Hampt}
\end{equation}

\begin{equation}
\nu' = \lambda' + 2{e^{\lambda}-1\over r} 
		-  2 r  e^{\lambda}(\psi_1^2+\psi_2^2) 
\label{slice}
\end{equation}
where a prime ($'$) denotes $\partial/\partial r$ and an overdot ($\dot{\ }$) 
denotes $\partial/\partial t$.

The Klein Gordon equation yields:

\begin{equation}
  \psi_1'' + \left( {2\over r} + {\nu'-\lambda'\over 2}  \right) \psi_1'
	+ e^\lambda\left( e^{-\nu}\omega^2-1  \right)\psi_1 
			- e^{\lambda-\nu}\ddot{\psi}_1
	+ e^{\lambda-\nu}{\dot{\nu}-\dot{\lambda}\over 2}(\dot{\psi}_1 
			  + \omega\psi_2)
	- 2 e^{\lambda-\nu}\omega\dot{\psi}_2 = 0
\label{kg1}
\end{equation}
and
\begin{equation}
  \psi_2'' + \left( {2\over r} + {\nu'-\lambda'\over 2}  \right) \psi_2'
	+ e^\lambda\left( e^{-\nu}\omega^2-1  \right)\psi_2 
				- e^{\lambda-\nu}\ddot{\psi}_2
	+ e^{\lambda-\nu}{\dot{\nu}-\dot{\lambda}\over 2}(\dot{\psi}_2 
				 - \omega\psi_1)
	+ 2 e^{\lambda-\nu}\omega\dot{\psi}_1 = 0.
\label{psi2pp}
\end{equation}

Another equation we will find useful is 
$G^\theta_\theta = 8\pi G T^\theta_\theta$, which evaluates to
\begin{eqnarray}
\lefteqn{   e^{-\lambda}\left( {\nu'-\lambda'\over 2r}  +  {1\over 2}\nu'' 
			  + {1\over 4}\nu'^2 - {1\over 4}\nu'\lambda'  \right)
 -  e^{-\nu}\left( {1\over 2}\ddot{\lambda} + {1\over 4}\dot{\lambda}^2 
		 - {1\over 4}\dot{\nu}\dot{\lambda} \right)  \nonumber } \\
& = & 
		e^{-\nu}\left( \dot{\phi}_1^2 + \dot{\phi}_2^2 + 
		  2\omega(\dot{\phi}_1\psi_2-\dot{\phi}_2\psi_1) 
					+ \omega^2(\psi_1^2+\psi_2^2) \right) 
  -  e^{-\lambda}(\psi_1'^2+\psi_2'^2) - (\psi_1^2+\psi_2^2).
\label{gthth}
\end{eqnarray}

We use equations (\ref{Hampt}) through (\ref{kg1}) to obtain the equilibrium 
solutions, by setting
\begin{eqnarray}
	\lambda(t,r) &=& \lambda_0(r)  \\
	\nu(t,r) &=& \nu_0(r)  \\
	\psi_1(t,r) &=& \phi_0(r)  \\
	\psi_2(t,r) &=& 0.
\end{eqnarray}

The equilibrium equations are then given by:
\begin{equation} \lambda_0' = {1 - e^{\lambda_0} \over r}
				+  r \left[
					 e^{\lambda_0}  ( \omega^2 e^{-\nu_0} + 1 )  \phi_0^2
				  + \phi_0'^2
				  \right]
\label{Ham0}
\end{equation}

\begin{equation} \nu_0' = {e^{\lambda_0} -1 \over r}
			  +  r \left[
					e^{\lambda_0}  ( \omega^2 e^{-\nu_0} - 1 )   \phi_0^2
					+ \phi_0'^2
				\right]
\end{equation}
 
\begin{equation} 
 \phi_0'' = - \left( {2\over r} + {\nu_0' - \lambda_0' \over 2} \right) \phi_0'
				 - e^{\lambda_0}   ( \omega^2 e^{-\nu_0}  - 1 )  \phi_0.
\end{equation}
We now introduce four perturbation 
fields---$\delta\lambda(t,r)$,
$\delta\nu(t,r)$,
$\delta\psi_1(t,r)$
and
$\delta\psi_2(t,r)$---and expand about the equilibrium configuration by
 writing:
\begin{eqnarray}
	\lambda(t,r) &=& \lambda_0(r)  + \delta\lambda(t,r)\\
	\nu(t,r) &=& \nu_0(r)+ \delta\nu(t,r)  \\
	\psi_1(t,r) &=& \phi_0(r) ( 1 + \delta\psi_1(t,r) )  \\
	\psi_2(t,r) &=& \phi_0(r) \delta\psi_2(t,r).
\end{eqnarray}

These last expressions are substituted into (\ref{Hampt}), (\ref{slice}), 
(\ref{kg1}) and (\ref{gthth}) to obtain the following equations for the 
perturbed quantities:

\begin{eqnarray}
(re^{-\lambda_0}\delta\lambda)' &=&  r^2 \left[ 
	 2\phi_0^2\delta\psi_1 - e^{-\nu_0}\omega^2\phi_0^2\delta\nu 
	+ 2e^{-\nu_0}\omega^2\phi_0^2\delta\psi_1 \right. \nonumber \\
& &\left.  - 2e^{-\nu_0}\omega\phi_0^2\delta\dot{\psi}_2
	+ 2e^{-\lambda_0}\phi_0'(\phi_0'\delta\psi_1 + \phi_0\delta\psi_1') 
	- e^{-\lambda_0}\phi_0'^2\delta\lambda
	\right]
\label{dlp}
\end{eqnarray}

\begin{equation}
 \delta\nu'-\delta\lambda' = \left(\nu_0'-\lambda_0' 
				 + {2\over r}\right) \delta\lambda
	- 4 r e^{\lambda_0}\phi_0^2\delta\psi_1
\label{dnp}
\end{equation}

\begin{eqnarray}
\delta\psi_1'' & + & \left( {2\over r} + {\nu_0'-\lambda_0'\over 2} + 
	 2{\phi_0'\over\phi_0}\right)\delta\psi_1'
 + {\phi_0'\over\phi_0}\left({\delta\nu'-\delta\lambda'\over 2}\right) 
			\nonumber \\
 & + & e^{\lambda_0}\left( \omega^2 e^{-\nu_0}-1  \right)\delta\lambda
 - e^{\lambda_0-\nu_0}\omega^2\delta\nu 
		 - e^{\lambda_0-\nu_0}\delta\ddot{\psi}_1
 - 2e^{\lambda_0-\nu_0}\omega\delta\dot{\psi}_2 = 0
\label{df1pp1}
\end{eqnarray}

\begin{eqnarray}
 &-&\delta\lambda e^{-\lambda_0}\left( {\nu_0'-\lambda_0'\over 2r} 
 +{1\over 2}\nu_0'' \right. 
	  + \left. {1\over 4}\nu_0'^2 - {1\over 4}\nu_0'\lambda_0' \right) 
				  \nonumber \\
  &+&  e^{-\lambda_0}\left( {\delta\nu'-\delta\lambda'\over 2r} 
				+ {1\over 2}\delta\nu''
  + {1\over 2}\nu_0'\delta\nu' - {1\over 4}\nu_0'\delta\lambda'
  - {1\over 4}\lambda_0'\delta\nu' \right) 
 - {1\over 2}e^{-\nu_0}\delta\ddot{\lambda} \nonumber \\
 = &-&\left[
	e^{-\nu_0}\omega^2\phi_0^2\delta\nu 
			 - e^{-\nu_0}\left( -2\omega\phi_0^2\delta\dot{\psi}_2 
		 + 2\omega^2\phi_0^2\delta\psi_1 \right)
  - e^{-\lambda_0}\phi_0'^2\delta\lambda \right. \nonumber \\
  &+&   \left. e^{-\lambda_0}\left( 2\phi_0'^2\delta\psi_1 
					+ 2\phi_0\phi_0'\delta\psi_1'  \right)
	+ 2\phi_0^2\delta\psi_1
\right].
\label{G_thth_pt}
\end{eqnarray}

The four equations above can be manipulated such that two variables, $\delta\nu
$ 
and $\delta\psi_2$ are eliminated, leaving us with only two equations in two 
unknowns.
To obtain the first of these two equations, we subtract 
(\ref{dlp}) from (\ref{df1pp1}) to get

\begin{eqnarray}  
\delta\psi_1'' = &-&\left( {2\over r} + {\nu_0'-\lambda_0' \over 2} \right) 
			\delta\psi_1'
-{\delta\lambda' \over  r\phi_0^2} 
			+ e^{\lambda_0-\nu_0}\delta\ddot{\psi}_1  \nonumber \\
&-&\left[ {\phi_0'\over \phi_0}\left( {\nu_0'-\lambda_0' \over 2} 
					+ {1\over r} \right)
		  + \left({\phi_0'\over \phi_0}\right)^2 +
		  {1- r\lambda_0' \over  r^2 \phi_0^2} + 
		  e^{\lambda_0-\nu_0}\omega^2 - e^{\lambda_0} \right]\delta\lambda   
 \nonumber \\
&+& 2 e^{\lambda_0} \left[
	 1 + e^{-\nu_0}\omega^2 
				 + e^{-\lambda_0}\left({\phi_0'\over \phi_0}\right)^2
	 +  r \phi_0\phi_0'
\right] \delta\psi_1.
\label{dpsipp}
\end{eqnarray}

To obtain the other equation, we differentiate (\ref{dnp}) with respect
to $r$, and substitute the resulting expression,  along with
(\ref{dlp}) and (\ref{dnp}), into (\ref{G_thth_pt}) to get
\begin{eqnarray}  
	\delta\lambda'' = &-&{3\over 2}(\nu_0' - \lambda_0') \delta\lambda'
	+\left[ 4 \phi_0'^2 + \lambda_0'' + {2\over r^2} 
			 - {(\nu_0' - \lambda_0')^2\over 2}
			- {2\nu_0' + \lambda_0' \over r} \right] \delta\lambda  
  + e^{\lambda_0-\nu_0}\delta\ddot{\lambda} \nonumber \\
 &-& 4 (2\phi_0\phi_0' - r  e^{\lambda_0}\phi_0^2) \delta\psi_1'  
				  \nonumber \\
 &-& 4\left[ 2\phi_0'^2 - 
		r e^{\lambda_0}\phi_0^2 
			 \left( 2 {\phi_0'\over \phi_0} + {2\nu_0' + \lambda_0' \over 2}
			  \right) \right] \delta\psi_1 ,
\label{dlpp}
\end{eqnarray}

where, differentiating (\ref{Ham0}) with respect to $r$ we have 
\begin{eqnarray}
 \lambda_0''& =& {e^{\lambda_0}-1\over r^2} - {e^{\lambda_0}\lambda_0'\over r}
	 +  \left[ e^{\lambda_0}  ( \omega^2 e^{-\nu_0} + 1 )  
				\phi_0^2 + \phi_0'^2 \right] \nonumber \\
	& + &  r \left[ -\nu_0'\omega^2 e^{\lambda_0-\nu_0} \phi_0^2
		+ e^{\lambda_0}(\omega^2 e^{-\nu_0} + 1)
				\left(\lambda_0'\phi_0^2 + 2{\phi_0\phi_0'}\right) 
		 + 2{\phi_0'\phi_0''} \right].
\label{df1pp}
\end{eqnarray}
(Note that (\ref{dpsipp}) omits a factor of $\exp(\lambda_0)$ which one
finds in the $\sim\delta\lambda/(r^2\phi_0^2)$ term of equation (34) in
\cite{GW}.) For the stability analysis, we assume a harmonic time
dependence, {\it i.e.},

\begin{eqnarray}
 \delta\psi_1(t,r) = \delta\psi_1(r)e^{i\sigma t} \nonumber \\
 \delta\lambda(t,r) = \delta\lambda(r)e^{i\sigma t}. \nonumber
\end{eqnarray}
Note that (\ref{dlpp}) and (\ref{df1pp}) contain only second derivatives with 
respect to time. There are good arguments for assuming $\sigma^2$ is purely 
real \cite{Jetzer,GW}, so we can determine instability by simply looking for
instances where $\sigma^2 < 0$.

As a further consideration, we note that the boson star system
admits a conserved Noether current, 
\begin{equation}
  J^\mu = {i\over 8\pi}g^{\mu\nu}(\phi\partial_\nu\phi^* - 
							  \phi^*\partial_\nu\phi),
\end{equation}
for which the corresponding charge or ``particle number'' is
\begin{eqnarray}
	  N &=& \int d^3 x \sqrt{-g}J^t \nonumber \\
		 &=& \int_0^\infty dr r^2 e^{(\lambda-\nu)/2}
		 \left(\dot{\psi}_1\psi_2 - \dot{\psi}_2\psi_1
			  + \omega(\psi_1^2 + \psi_2^2) \right) .
\end{eqnarray}

Conventional stability analysis (see, {\it e.g.}, \cite{ST}) demands
that we consider only perturbations for which the total charge is
conserved.  Thus we compute the variation in the charge, $\delta N$,
and work to ensure $\delta N = 0$.
In practice, since we cut off the grid at finite radius, it makes
sense to consider the function $\delta N(r)$, the total charge enclosed
in a sphere with surface area $4\pi r^2$.  This quantity is

\begin{eqnarray}
	\delta N(r) &=& {1\over \omega} \int_0^r 
d\tilde{r}\, \tilde{r}^2 e^{(\nu_0-\lambda_0)/2} \phi_0^2 
  \nonumber \\
  & \times & \left\{   {\delta\lambda'\over 2 \tilde{r}\phi_0^2}
		+ {1\over 2}\left[ e^{\lambda_0-\nu_0}\omega^2 + \left({\phi_0'\over
 \phi_0}\right)^2 +
		{1- \tilde{r}\lambda_0' \over  \tilde{r}^2 \phi_0^2}  \right]\delta\lambda
	 \right. \nonumber \\
 &-&\left.  {\phi_0'\over\phi_0}\delta\psi_1'
	 -\left[ - e^{\lambda_0-\nu_0}\omega^2 + \left({\phi_0'\over \phi_0}\right)^2 
+ e^{\lambda_0}
	  \right]  \delta\psi_1 \right\},
\label{deltaN}
\end{eqnarray}
where primes denote $\partial/\partial \tilde{r}$.  (Note that (\ref{deltaN})
contains a term involving $\delta\psi_1'$, which was not included in
equation (35) of \cite{GW}.)  We then demand that
$\delta N \rightarrow 0$ as $r\rightarrow\infty$.

The boundary conditions are as follows:

At $r=0$:
\begin{eqnarray}
  \lambda_0  &=&  0    \nonumber  \\
  \nu_0  &=&   0    \nonumber  \\
  \phi_0'  &=&  0     \nonumber  \\
 \phi_0''  &=&  - {1\over 3}( \omega^2 - 1) \phi_0   \\
 \delta\psi_1''  &=&  {1\over 3}\left[  -{3\delta\lambda''\over 2 \phi_0^2}    
	  + \left( 2(\omega^2 + 1) - \sigma^2 \right)\delta\psi_1 \right]  \\
 \delta\lambda  &=&  0       \nonumber  \\
 \delta\lambda'  &=&  0.       \nonumber 
\end{eqnarray}
As $r\rightarrow\infty$:
\begin{eqnarray}
	& \delta N   \rightarrow   0 & \nonumber  \\
& \delta\psi_1  \rightarrow   0 & \nonumber  \\
& \delta\lambda \rightarrow   0. & \nonumber  
\end{eqnarray}

To solve the system (\ref{dlpp}) and (\ref{df1pp}) subject to the above
boundary conditions, for a given value of $\phi_0(0)$, we resort to the
method of ``shooting," first for the equilibrium solutions, then for the
perturbed quantities.  Specifically, we choose a value for $\omega$ and solve
 the
equilibrium equations numerically by integrating outward from $r=0$.
We do this repeatedly, performing a ``binary search" on $\omega$ (as
described in \cite{RB}) until the boundary conditions for the
equilibrium quantities are satisfied.

Due to the linearity of the problem, we can choose $\delta\psi_1(0)$
arbitrarily.  We then have two parameters left, namely $\sigma^2$ and
$\delta\lambda''(0)$.  To make matters easy at first, we consider
perturbations very close to the transition between stability and
instability.  At the transition point, $\sigma^2$ is zero.  Thus for
boson stars near the transition point, we choose $\sigma^2=0$ and shoot
on the parameter $\delta\lambda''(0)$ until the boundary conditions are
satisfied.  As Gleiser and Watkins \cite{GW} note, the transition point
occurs at the maximum boson star mass; so we can take two slightly
different equilibrium solutions near the maximum mass and subtract them
to generate solutions which should agree with those obtained from the
perturbation problem.  We use this method to obtain a trial value of
$\delta\lambda''(0)$, and also as a way of checking the final solution
we obtain from the perturbation analysis.

For more general configurations ($\sigma^2 \neq 0$), we choose a value
of $\sigma^2$ and shoot on $\delta\lambda''(0)$ until we find $\delta
N$ at the outer boundary of the grid to be less than some tolerance
value.  Then we use the fact (gleaned from experience) that if
$\sigma^2$ is too large (too positive), $\delta N$ will have a local
minimum, the value of which will be less than zero  ({\it i.e.},
$\delta N(r)$ will dip below zero and then turn back up at larger
radii).  If $\sigma^2$ is too low there will be no such local minimum.  We
use these two criteria to select the value of $\sigma^2$ via a binary
search.  
Thus our two-dimensional eigenvalue-finding algorithm consists simply
of two (nested) binary searches, one in each direction:
For each value of $\sigma^2$ tried, a full binary search on
the parameter $\delta\lambda''(0)$ is performed to drive $\delta
N(r_{\tt max})\rightarrow 0$.  Then the solution of $\delta N(r)$ is
examined for the behavior described above, and a new value of
$\sigma^2$ is selected, and so on until both $\delta\lambda''(0)$ and
$\sigma^2$ have been found to some desired precision.

\end{document}